%% file: double.tex
\documentclass[journal]{IEEEtran}

\usepackage{cite}
\usepackage{graphicx}  
\usepackage{amsmath}   
\usepackage{subfigure}
\usepackage{algorithm}
\usepackage{algorithmic}
\usepackage{amssymb}
\usepackage{color}
\usepackage{pbox}
\usepackage{cases}
\usepackage{multirow}
\usepackage{float}
\DeclareMathOperator*{\st}{s.t.}
\newcommand{\trans}{^{T}}

\input{commands}

\allowdisplaybreaks

\DeclareMathOperator*{\argmin}{arg\,min}

\begin{document}

\title{Distributed Optimization for Coordinated Beamforming in Multi-Cell Multigroup Multicast Systems: Power Minimization and SINR Balancing}

\author{Oskari Tervo,~\IEEEmembership{Student~Member,~IEEE}, Harri Pennanen,~\IEEEmembership{Member,~IEEE}, Dimitrios Christopoulos,~\IEEEmembership{Member,~IEEE}, Symeon Chatzinotas,~\IEEEmembership{Senior Member,~IEEE}, and Bj\"orn Ottersten,~\IEEEmembership{Fellow,~IEEE}
\thanks{This work was partially supported by the National Research Fund, Luxembourg, under the projects $\mathrm{SATSENT}$, $\mathrm{PROSAT, ECLECTIC}$, and $\mathrm{INWIPNET}$, and by the European Commission, H2020, under the project $\mathrm{SANSA}$. The research was also supported in part by Infotech Oulu Doctoral
Program and the Academy of Finland under project Message and CSI Sharing for Cellular Interference Management with Backhaul Constraints ($\mathrm{MESIC}$)
belonging to the WiFIUS program with NSF, and WiConIE. The first author has been supported by Oulu University Scholarship Foundation, Nokia Foundation, Tauno T{\"o}nning Foundation, Walter Ahlstr{\"o}m Foundation, and KAUTE foundation. Parts of this paper have been presented at the IEEE International Conference
on Communications, Kuala Lumpur, Malaysia,
May 2016.}
\thanks{O. Tervo and H. Pennanen are with Centre for Wireless Communications, University of Oulu,
Finland. Email: \{oskari.tervo,harri.pennanen\}@oulu.fi.}
\thanks{D. Christopoulos is with Newtec Cy., Belgium. Email: dchr@newtec.eu.}
\thanks{S. Chatzinotas and B. Ottersten are with the Interdisciplinary Centre for Security, Reliability and Trust, University of Luxembourg, Luxembourg. Email: \{symeon.chatzinotas, bjorn.ottersten\}@uni.lu}}

%

\maketitle

\begin{abstract}
This paper considers coordinated multicast beamforming in a multi-cell multigroup multiple-input single-output system. Each base station (BS) serves multiple groups of users by forming a single beam with common information per group. We propose centralized and distributed beamforming algorithms for two different optimization targets. The first objective is to minimize the total transmission power of all the BSs while guaranteeing the user-specific minimum quality-of-service targets. The semidefinite relaxation (SDR) method is used to approximate the non-convex multicast problem as a semidefinite program (SDP), which is solvable via centralized processing. Subsequently, two alternative distributed methods are proposed. The first approach turns the SDP into a two-level optimization via primal decomposition. At the higher level, inter-cell interference powers are optimized for fixed beamformers while the lower level locally optimizes the beamformers by minimizing BS-specific transmit powers for the given inter-cell interference constraints. The second distributed solution is enabled via an alternating direction method of multipliers, where the inter-cell interference optimization is divided into a local and a global optimization by forcing the equality via consistency constraints. We further propose a centralized and a simple distributed beamforming design for the signal-to-interference-plus-noise ratio (SINR) balancing problem in which the minimum SINR among the users is maximized with given per-BS power constraints. This problem is solved via the bisection method as a series of SDP feasibility problems. The simulation results show the superiority of the proposed coordinated beamforming algorithms over traditional non-coordinated transmission schemes, and illustrate the fast convergence of the distributed methods.
\end{abstract}


\section{Introduction}
\label{sec:intro}
The performance of a communication system can be improved with a signal processing technique called transmit beamforming (or equivalently precoding). Given a certain quality of channel state information, properly designed advanced multi-antenna beamforming techniques have a potential of significantly improving spectral efficiency. However, inter-cell interference (ICI) may be a limiting performance factor without proper interference coordination between neighboring cells. To address this issue, coordinated beamforming has been adopted in the literature as a powerful interference management method. The idea of the approach is to involve the inter-cell interference in the beamforming optimization design, which improves the system performance especially at cell-edge areas \cite{Gesbert-10}.
In coordinated beamforming, each data stream is linearly precoded in the spatial domain and transmitted from a single base station (BS). More specifically, the aim is to achieve a network performance target with some user-specific quality of service constraints, by jointly designing precoded data transmissions
among BSs.
Coordinated beamforming schemes rely on the availability of channel state information (CSI) at the BSs. Depending on the implementation, the total amount of required CSI is different.
In general, a centralized algorithm refers to a case where global CSI is required, which essentially means the knowledge of the channels between all the BSs and all the users in the system. On the other hand, in the distributed methods, each BS has to know only its channels towards all the users in the coordinating network to perform the optimization. In the literature, this is  usually called local CSI. Throughout this paper, the acquired global or local CSI is assumed to be perfect.
Generally, decentralized schemes may be easier to implement in practice since they reduce the signaling overhead and require lower computational requirements per processing unit.
In the recent wireless communications research, coordinated beamforming has received an extensive attention for various system design objectives, such as sum power minimization \cite{Rashid-Farrokhi-98a}, minimum SINR maximization (also called SINR balancing problem in the literature) \cite{Huang-12}, sum rate maximization \cite{Shi-11} and energy efficiency maximization \cite{TervoWsumEE-15, Tervo-16Arxiv}.
The sum power minimization problem \cite{Rashid-Farrokhi-98a} is to minimize the sum power of all the BSs with certain user-specific SINR targets.
This system design objective is of practical interest for wireless applications which have stringent data rate and delay constraints. For the sum power minimization problem, the works of \cite{Rashid-Farrokhi-98a,Bengtsson-99,Bengtsson-01,Wiesel-06} and \cite{Rashid-Farrokhi-98a,Dahrouj-10,Tolli-11,Pennanen-11,Pennanen-14a,Shen-12} have proposed centralized and distributed beamforming designs, respectively. The dual and primal decomposition was applied in \cite{Tolli-11} and \cite{Pennanen-11}, respectively, while the work of \cite{Shen-12} proposed a robust alternating direction method of multipliers (ADMM)-based approach. On the other hand, the problem of SINR balancing targets at maximizing the minimum SINR among the users. In this objective, the available power is used to provide the best possible Quality of Service (QoS) for the users in a fair manner. Coordinated beamforming for this optimization target has been studied, e.g., in \cite{He-12MaxMin, Cai-12, Hou-14}.

\subsubsection*{Related Work}
The previously presented literature assumes an independent data stream for each user, which is known as unicast beamforming. Another interesting technique which has also been recognized in LTE standards is called multicast beamforming, where one stream is intended to a group of users.
It has a great potential to address the nature of future traffic demands, e.g., to support demanding video broadcasting applications. Furthermore, physical layer multicasting has applications in satellite communications \cite{Vazquez-16,Christopoulos-16Sat}. For the power minimization target, a physical layer multicasting problem was originally proposed in \cite{Sidiropoulos-06}, proven NP-hard and accurately approximated by the semidefinite relaxation (SDR) and Gaussian randomization techniques. Bornhorst {\it et al.} \cite{Bornhorst-12} proposed a distributed successive convex approximation method to solve the multicasting problem in relay networks.
In \cite{Karipidis-08}, a unified framework was derived for physical layer multigroup multicasting, where independent sets of common data are transmitted to different interfering groups of users.
They proved the NP-hardness of the sum power minimization and the minimum SINR maximization problems in a multicast multigroup system with a sum power constraint (also known as the QoS and the max-min fair problems), and proposed accurate approximations to solve the problems.
The works of \cite{Christopoulos-14,Christopoulos-14b} derived a consolidated solution for the weighted max-min fair multigroup multicast beamforming under per-antenna power constraints.
This work was extended for the sum rate maximization problem in \cite{Christopoulos-15}, where the important aspect of user scheduling was also considered.
The coordinated multicast beamforming problem with a single group per cell was considered in \cite{Xiang-13}, where the authors derived a distributed and centralized method for the QoS and max-min fair problems, respectively. In \cite{He-15}, the authors proposed a centralized algorithm for the energy efficiency maximization problem in a multi-cell multigroup system.
In the literature, however, there is a lack of generic centralized and distributed algorithms for the sum power minimization and SINR balancing problems in a multi-cell multicast system with multiple groups per cell.
\subsubsection*{Contributions}
In this paper, we propose centralized and distributed beamforming designs for multi-cell multigroup multicast systems. Two different optimization targets are considered, i.e., power minimization and SINR balancing. We first generalize the methods of \cite{Karipidis-08} from single-cell to a multi-cell system model (i.e., the centralized methods) and \cite{Xiang-13} to a multigroup system model (i.e., centralized and primal decomposition -based distributed method for sum power minimization and a centralized method to SINR balancing problem). To be more specific, in the first design, we aim at minimizing the total transmission power of the system while guaranteeing predefined minimum SINRs for all the active users. The resulting problem is non-convex and difficult to tackle as such. We propose to use SDR to approximate the problem as a semidefinite program (SDP) \cite{Karipidis-08,Xiang-13}. If the SDR results in solutions where all the covariance matrices are rank-1, then the algorithms are optimal for the original problem. Otherwise, the Gaussian randomization method is utilized to provide a sub-optimal, but feasible, beamforming solution. However, the simple power scaling as in \cite{Xiang-13} cannot be applied to a multi-group setting due to multiple power variables in each cell. Thus, differently from \cite{Xiang-13}, we propose a linear programming method to perform Gaussian randomization.
We further propose two alternative methods to obtain a distributed beamforming implementation. Firstly, we generalize the primal decomposition method of \cite{Xiang-13} to a multigroup system model by reformulating the one-level SDP into two optimization levels. In the higher level, upper bounding levels for inter-cell interference powers are optimized while the lower level is in charge of optimizing the beamformers for a given set of interference power constraints. Secondly, differently
from \cite{Xiang-13}, we propose another distributed solution via an alternating direction method of multipliers, where the inter-cell interference optimization is divided into local and global optimization by forcing the equality via consistency constraints. The distributed methods require only local CSI at each BS and low-rate backhaul scalar information exchange.
In the second problem formulation, we consider a multicast SINR balancing problem to obtain fairness between the users. More specifically, the aim is to maximize the minimum SINR of all the users in the network. In the proposed centralized design, we again generalize the method of \cite{Karipidis-08} from a single-cell to
a multi-cell and \cite{Xiang-13} to a multigroup system model by reformulating
 the original nonconvex max-min fractional program based on SDR. Then, the problem is reformulated to find a bisection type algorithm where the SDP feasibility problem is solved in each iteration. Subsequently, differently from the earlier works, by limiting the maximum inter-cell interference towards each user at each BS, we propose a simple distributed beamforming design which can be implemented without any information exchange among the cells. The superiority of the proposed algorithms over the conventional transmission schemes is demonstrated via numerical examples. We also compare the convergence rates of the distributed methods and illustrate when the proposed algorithms are optimal, i.e., produce rank-1 solutions.

Parts of this paper have been published in the previous conference publication \cite{Pennanen-16}, i.e., the centralized approach and the primal decomposition based distributed implementation for the sum power minimization problem with preliminary simulation results. The following additional contributions can be found in this paper which have not been in the previous work. Here, we propose an alternative distributed beamforming design based on the ADMM method for the power minimization problem. We further provide a centralized and a distributed algorithm for the SINR balancing problem, and show how the Gaussian randomization is performed in this formulation. Finally, we have extended the simulation model to take into account varying interference conditions, i.e., we have added a cell separation parameter into simulations. Using this, we have provided a more extensive set of simulations to illustrate the effectiveness of the proposed methods. 
\subsubsection*{Organization and Notation} 
The paper is organized as follows. In Sections \ref{sec:System_Model} and \ref{sec:ProblemFormulation}, the multi-cell multigroup multicast system is introduced, and the optimization problems are formulated, respectively. In sections \ref{sec:CentralizedAlg} and \ref{sec:DecentralizedAlg}, the centralized and distributed beamforming algorithms for the power minimization problem are derived, respectively. Section \ref{sec:SINRbalancing} presents the centralized and distributed algorithm for the SINR balancing problem. The performance of the proposed algorithms is examined in Section \ref{sec:SimulationResults} via numerical examples. Finally, conclusions are drawn in Section \ref{sec:Conclusion}.

The following notations are used.
Bold face lower and upper case characters denote column vectors and matrices, respectively. The operators $\left(\cdot\right)\herm$ and $\mathrm{Tr}(\cdot)$ correspond to the conjugate transpose and the trace operator. Calligraphic letters are for sets and notation $\mathbf{X}\succeq0$ means that $\mathbf{X}$ is positive semidefinite.  
\section{System Model}
\label{sec:System_Model} 
The multi-cell multigroup multicasting system consists of $B$ BSs, $G$ groups and $U$ users. Each BS has $A$ transmit antennas, whereas each user is equipped with a single receive antenna. We denote the sets of BSs, groups and users by $\mathcal{B}=\{1,\ldots,B\}$, $\mathcal{G}=\{1,\ldots,G\}$ and $\mathcal{U}=\{1,\ldots,U\}$, respectively. Each group of users is served by an independent data stream transmitted from a single BS. Therefore, there exists inter-group interference between the groups of the same cell and between the groups belonging to different cells. The former interference type is known as intra-cell interference and the latter as inter-cell interference. The number of groups served by BS $b$ is given by $G_b$, and the corresponding set of groups is denoted by $\mathcal{G}_b$. The number of users in group $g$ is denoted by $U_g$, and the corresponding set of users is given by $\mathcal{U}_g$. Note that each user belongs only to a single group. Thus, the sets of users belonging to different groups are disjoint. This can be mathematically expressed as $\mathcal{U}_i \cap \mathcal{U}_j = \emptyset$, $\forall i,j \in \mathcal{G}, i \neq j$.
The signal received by user $u$ is given by
\begin{eqnarray} \label{eq:RxSignal}
{y}_{u} &=& \overbrace{{\vec h}_{b,u}\herm {\vec w}_{g} {s}_{g}}^\text{desired signal} + \overbrace{\sum\limits_{i \in \mathcal{G}_{b} \setminus \{g\}} {\vec h}_{b,u}\herm {\vec w}_{i} {s}_{i}}^\text{intra-cell interference} \nonumber \\
&& + \underbrace{\sum\limits_{j \in \mathcal{B} \setminus \{b\}} \sum\limits_{k \in \mathcal{G}_j} {\vec h}_{j,u}\herm {\vec w}_{k} {s}_{k}}_\text{inter-cell interference} + {n}_{u}, \nonumber \\
&& \forall b \in \mathcal{B}, \forall g \in \mathcal{G}_b, \forall u \in \mathcal{U}_g
\end{eqnarray}
where ${\vec h}_{b,u} \in \C^{A}$ is the channel vector from $b$th BS to $u$th user, ${\vec w}_{g} \in \C^{A}$ is the unnormalized beamformer of group $g$, ${s}_{g} \in \C$ is the corresponding normalized information symbol and ${n}_{u} \ \sim \mathcal{C} \mathcal{N} (0, \sigma_{u}^{2})$ is the complex realization of white Gaussian noise with zero mean and variance $\sigma_{u}^{2}$.

Gaussian codebook is assumed to be used for each data stream. Thus, the rate for user $u$ can be expressed as
\begin{eqnarray} \label{eq:UserRate}
\displaystyle r_{u} = \log_{2} \left(1 + \Gamma_{u}\right)
\end{eqnarray}
where $\Gamma_{u}$ denotes the SINR of user $u$. The mathematical expression of $\Gamma_{u}$ is written as
\begin{eqnarray} \label{eq:SINR}
\displaystyle \Gamma_{u} = \frac{|{\vec h}_{b,u}\herm {\vec w}_{g}|^{2}}{{\sigma_{u}^{2} + \sum\limits_{j \in \mathcal{B}} \sum\limits_{k \in \mathcal{G}_j \setminus \{g\}} |{\vec h}_{j,u}\herm {\vec w}_{k}|^{2}}}
\end{eqnarray}
where user $u$ belongs to a group $g$.
\section{Problem Formulation}
\label{sec:ProblemFormulation}
We consider two different optimization targets. In the first one, the (network-level) optimization goal is to minimize the sum transmission power of the BSs while satisfying the user-specific rate targets for active users. The mathematical expression of the problem is given by
\begin{equation} \label{eq:SPMinMulticast1}
\begin{array}{ll}
\displaystyle \underset{\{{\vec w}_{g}\}_{g \in \mathcal{G}}}{\mathrm{min.}}  &  \displaystyle  \sum\limits_{g \in \mathcal{G}} {\rm Tr} \left({\vec w}_{g} {\vec w}_{g}\herm \right)\\
{\mathrm{s.\ t.}}
& \displaystyle \log_{2} \left(1 + \Gamma_{u}\right) \geq R_{u}, \\
& \forall b \in \mathcal{B}, \forall g \in \mathcal{G}_b, \forall u \in \mathcal{U}_g
\end{array}
\end{equation}
where $R_{u}$ is a predefined rate target for user $u$. In the multiple-input single-output (MISO) problem setting in \eqref{eq:SPMinMulticast1}, there exists a direct mapping between the user-specific rate and SINR. Hence, the rate targets $\{R_{u}\}_{u \in \mathcal{U}}$ can be equivalently changed into SINR targets $\{\gamma_u\}_{u \in \mathcal{U}}$.
The resulting problem is given by
\begin{equation} \label{eq:SPMinMulticast}
\begin{array}{ll}
\displaystyle \underset{\{{\vec w}_{g}\}_{g \in \mathcal{G}}}{\mathrm{min.}}  &  \displaystyle  \sum\limits_{g \in \mathcal{G}} {\rm Tr} \left({\vec w}_{g} {\vec w}_{g}\herm \right)\\
{\mathrm{s.\ t.}}
& \displaystyle \frac{|{\vec h}_{b,u}\herm {\vec w}_{g}|^{2}}{{\sigma_{u}^{2} + \sum\limits_{j \in \mathcal{B}} \sum\limits_{k \in \mathcal{G}_j \setminus \{g\}} |{\vec h}_{j,u}\herm {\vec w}_{k}|^{2}}} \geq \gamma_{u}, \\
& \forall b \in \mathcal{B}, \forall g \in \mathcal{G}_b, \forall u \in \mathcal{U}_g
\end{array}
\end{equation}
where $\gamma_u=2^{R_u}-1$ is the fixed SINR target.
Note that \eqref{eq:SPMinMulticast} may be infeasible for some system settings and/or channel conditions. This can happen in interference limited scenarios, e.g., if the number of antennas at some BS $b$ is smaller than the total number of users in neighboring cells plus the sum of the user groups served by BS $b$. Nevertheless, admission control is responsible for handling infeasibility issues by alleviating the system-specific requirements such as reducing the number of simultaneously scheduled users or decreasing the minimum SINR values. \cite{Stridh-06}. In \cite{Wiesel-06} and \cite{Xiang-13}, the feasibility issue was considered for unicast and multicast beamforming concepts, respectively. A joint admission control and multicast beamforming problem was solved and discussed in \cite{Matskani-09}. The feasibility of \eqref{eq:SPMinMulticast} is assumed in the rest of this paper.

Problem \eqref{eq:SPMinMulticast} is difficult to solve as such because it is non-convex and NP-hard. In fact, a conceptually simple single-cell multicast problem was shown to be NP-hard in \cite{Sidiropoulos-06}, and \eqref{eq:SPMinMulticast} is a more generic multi-cell version of the problem in \cite{Sidiropoulos-06}. It is worth mentioning that in case of heterogeneous power requirements between the BSs, BS-specific priority weights could be easily included in problem \eqref{eq:SPMinMulticast}. For the ease of notation, however, they are left out in this paper. On the other hand, we could easily include different power constraints in \eqref{eq:SPMinMulticast}, e.g., BS-specific, antenna-specific or antenna group -specific power constraints. All of these constraints admit an SDP formulation and they are readily separable between the base stations. By adding the given power constraints will reduce the size of the feasible solution set of \eqref{eq:SPMinMulticast}.

The second problem of interest is a coordinated multicast SINR balancing problem where we aim at providing fairness among the users by maximizing the minimum SINR of the users, subject to BS-specific power constraints. The problem can be cast as
\begin{equation} \label{eq:SINRbalancingMulticast0}
\begin{array}{ll}
\displaystyle \underset{\{{\vec w}_{g}\}_{g \in \mathcal{G}}}{\mathrm{max.}} \underset{u\in\mathcal{U}}{\mathrm{min.}}  &  \displaystyle  \frac{|{\vec h}_{b,u}\herm {\vec w}_{g}|^{2}}{{\sigma_{u}^{2} + \sum\limits_{j \in \mathcal{B}} \sum\limits_{k \in \mathcal{G}_j \setminus \{g\}} |{\vec h}_{j,u}\herm {\vec w}_{k}|^{2}}}\\
{\mathrm{s.\ t.}}
& \displaystyle \sum_{g\in\mathcal{G}_b}{\rm Tr}(\mathbf{w}_g\mathbf{w}_g\herm) \leq P_b, \forall b \in \mathcal{B}
\end{array}
\end{equation}
where $P_b$ is a maximum transmit power allowed for BS $b$. Similarly to the power minimization problem, \eqref{eq:SINRbalancingMulticast0} is hard to tackle as such due to its non-convexity and NP-hardness. Note that we could add the priority weights for each user in problem \eqref{eq:SINRbalancingMulticast0} if some users are more important than others. However, the weights are left out for the ease of notation. The algorithms proposed in this paper can be straightforwardly generalized to different power constraints, e.g., for antenna-specific or antenna group -specific power constraints.

\section{Power Minimization Problem}
\label{sec:CentralizedAlg}
Here we aim at minimizing the total transmit power of all the BSs with a given user-specific SINR constraints. We first derive a centralized algorithm, which is followed by the primal decomposition -based and ADMM-based distributed methods, respectively. Subsequently, the distributed Gaussian randomization method is presented for the cases where the solution is not rank-1. Finally, practical aspects of the distributed methods are discussed, e.g., the effects of limiting the number of iterations and signaling overhead caused by the algorithms.
\subsection{Centralized Beamforming Design}
\label{sec:CentralizedAlg}
Due to the non-convexity of problem \eqref{eq:SPMinMulticast}, a more tractable formulation which can be solved efficiently is desired. To this end, the SDR method can be used to obtain a convex approximation of problem \eqref{eq:SPMinMulticast}. Specifically, the term
${\vec w}_{g}{\vec w}_{g}\herm$ is replaced with a semidefinite matrix ${\vec W}_{g}$, $\forall g \in \mathcal{G}$.
The resulting convex SDP can be stated as
\begin{IEEEeqnarray}{ll}\label{eq:SPMinMulticastApproximated}
\displaystyle \underset{\{{\vec W}_{g}\}_{g \in \mathcal{G}}}{\mathrm{min.}}  &  \displaystyle  \sum\limits_{g \in \mathcal{G}} {\rm Tr} \left({\vec W}_{g}\right) \IEEEyessubnumber\\
{\mathrm{s.\ t.}}
& \displaystyle \frac{{\rm Tr} \left({\vec H}_{b,u}{\vec W}_{g}\right)}{{\sigma_{u}^{2} + \sum\limits_{j \in \mathcal{B}} \sum\limits_{k \in \mathcal{G}_j \setminus \{g\}} {\rm Tr} \left({\vec H}_{j,u} {\vec W}_{k} \right)}} \geq \gamma_{u}, \IEEEyessubnumber \label{eq:SPMinMulticastApproximated:SINR} \\
& \forall b \in \mathcal{B}, \forall g \in \mathcal{G}_b, \forall u \in \mathcal{U}_g \IEEEyessubnumber \\
& {\vec W}_{g} \succeq 0, \forall g \in \mathcal{G} \IEEEyessubnumber
\end{IEEEeqnarray}
where ${\vec H}_{b,u}={\vec h}_{b,u}{\vec h}_{b,u}\herm$. In order to solve \eqref{eq:SPMinMulticastApproximated}, global CSI is required at a central controlling unit or at each BS. If all the optimal covariance matrices produced by \eqref{eq:SPMinMulticastApproximated} are rank-1, the solution is also optimal for the original problem \eqref{eq:SPMinMulticast}. In this case, the eigenvalue decomposition can be used to extract the optimal beamformers $\{{\vec w}_{g}^{*}\}_{g \in \mathcal{G}}$ from $\{{\vec W}_{g}^{*}\}_{g \in \mathcal{G}}$. The resulting beamformers are given by ${\vec w}_{g}^{*}=\sqrt{\lambda_{g}}{\vec u}_{g}$, $\forall g \in \mathcal{G}$,
where $\lambda_{g}$ and ${\vec u}_{g}$ are the principal eigenvalue and eigenvector of ${\vec W}_{g}^{*}$. However, due to the relaxation, the rank of ${\vec W}_{g}$ can be larger than one. Consequently, an optimal solution of \eqref{eq:SPMinMulticastApproximated} is not necessarily optimal for \eqref{eq:SPMinMulticast}.

For some specific problems in the literature the SDR provides optimum solutions, e.g., the optimal unicast beamforming in \cite{Bengtsson-01}.
Generally, however, the unit rank solution of the relaxed problem cannot be guaranteed due to the NP-hardness of the multicast problem. To this end, a general approach is to apply a rank-one approximation over the higher rank solution.
In case of multicast beamforming, the Gaussian randomization method has been shown to result in the highest accuracy \cite{Luo-10}.
Let us denote by $\{{\vec W}_{g}^{*}\}_{g \in \mathcal{G}}$ the optimal symmetric positive semidefinite matrices of the relaxed problem. In the Gaussian randomization,
we generate a rank-one beamforming approximation of the original problem as a complex Gaussian vector with zero mean and covariance equal to ${\vec W}_{g}^{*}$, i.e.  $\hat{{\vec w}}_g \ \sim \mathcal{C} \mathcal{N}(0, {\vec W}_{g}^{*} )$, $\forall g \in \mathcal{G}$. Vector $\hat{{\vec w}}_g$ is called a candidate beamforming solution. For the candidate beamformers, however, the feasibility of the solution to the original problem is not guaranteed. Thus, the feasibility of each candidate beamformer has to be checked by solving a power minimization problem. The aim is to minimize the transmit powers $\{p_g\}_{g \in \mathcal{G}}$ with the user-specific SINR targets $\{\gamma_u\}_{u \in \mathcal{U}}$, for each fixed candidate beamformer $\{\hat{{\vec w}}_{g}\}_{g \in \mathcal{G}}$ (normalized to have unit power). This yields a linear program (LP)
\begin{equation} \label{eq:GR_PowOpt_Centr}
\begin{array}{ll}
\displaystyle \underset{\{p_{g}\}_{g \in \mathcal{G}}}{\mathrm{min.}}  &  \displaystyle  \sum\limits_{g \in \mathcal{G}} p_{g}\\
{\mathrm{s.\ t.}}
& \displaystyle \frac{p_g \left|{\vec h}_{b,u}\hat{{\vec w}}_{g}\right|^{2}}{{\sigma_{u}^{2} + \sum\limits_{j \in \mathcal{B}} \sum\limits_{k \in \mathcal{G}_j \setminus \{g\}} p_k \left|{\vec h}_{j,u}\hat{{\vec w}}_{k}\right|^{2}}} \geq \gamma_{u}, \\
& \forall b \in \mathcal{B}, \forall g \in \mathcal{G}_b, \forall u \in \mathcal{U}_g.
\end{array}
\end{equation}
After solving \eqref{eq:GR_PowOpt_Centr}, the beamforming solution for each set of candidate beamformers is obtained as ${\vec w}_{g}=\sqrt{p_{g}^{*}} \hat{{\vec w}}_{g}$, $\forall g \in \mathcal{G}$, where $p_{g}^{*}$ is the optimal power associated with fixed candidate beamformer $\hat{{\vec w}}_{g}$.
Finally, after generating a predetermined number of candidate solutions, the one that yields the lowest objective value of the original problem is chosen. Note that the beamformers $\{{\vec w}_{g}\}_{g \in \mathcal{G}}$ are sub-optimal, but feasible, for the original problem.
The accuracy of this approximate solution is measured by the distance of the approximate objective value and the optimal value of the relaxed problem. This accuracy increases with the increasing number of Gaussian randomizations.
The proposed centralized multicast approach is summarized in {\it Algorithm~\ref{alg:SPMinMulticastAlgCentr}}. The algorithm requires the knowledge of the global CSI either in the central controlling unit (if the controller performs the optimization) or in each BS $b$ (if the CSI is shared between the BSs). In the latter case, the BSs run {\it Algorithm~\ref{alg:SPMinMulticastAlgCentr}} in parallel, i.e., at the same time.
 It is worth mentioning that an alternative iterative solution based on iterative convex approximation without introducing the SDR was proposed in \cite{Bornhorst-11}. However, the advantages of the SDR approach lie in the fact that only a single problem needs to be solved if the solution is rank-1, and even the Gaussian randomization requires solving a linear program which can be implemented efficiently. The method of \cite{Bornhorst-11} can be more useful in the cases where the ranks of the covariance matrices would be far from rank-1.

\begin{algorithm} [tbp!]
\caption{Centralized multicast beamforming for power minimization}
\label{alg:SPMinMulticastAlgCentr}
\begin{algorithmic}[1]
\STATE Compute optimal transmit covariance matrices $\{{\vec W}_{g}^{*}\}_{g \in \mathcal{G}}$ by solving the SDP \eqref{eq:SPMinMulticastApproximated}.
\STATE Check whether the ranks of $\{{\vec W}_{g}^{*}\}_{g \in \mathcal{G}}$ are all one or not. If the ranks are one, apply eigenvalue decomposition for $\{{\vec W}_{g}^{*}\}_{g \in \mathcal{G}}$ to find optimal beamformers $\{{\vec w}_{g}^{*}\}_{g \in \mathcal{G}}$ for the original problem. Otherwise, apply Gaussian randomization with power optimization \eqref{eq:GR_PowOpt_Centr} to find feasible, but sub-optimal, beamformers $\{{\vec w}_{g}\}_{g \in \mathcal{G}}$.
\end{algorithmic}
\end{algorithm}

\subsection{Distributed Beamforming Designs}
\label{sec:DecentralizedAlg}

The algorithm derived in the previous section requires centralized processing due to the coupling ICI terms in \eqref{eq:SPMinMulticastApproximated:SINR}. In this section, we derive distributed methods which decouple the problems so that each BS can perform beamformer optimization based on local channel state information and low-rate scalar information exchange.
Two alternative formulations are proposed, one based on primal decomposition and the other based on alternating direction method of multipliers. We derive the algorithms, explain the distributed implementations, summarize the details in step-by-step algorithms and discuss some practical aspects. The Gaussian randomization method is presented in case the solution is not rank-1.

To apply distributed optimization, we first reformulate \eqref{eq:SPMinMulticastApproximated}. In this respect, we separate interference power to intra-cell and inter-cell terms, and add auxiliary variables to denote the ICI terms.
Now, the coupling is transferred from beamformers to ICI variables.
The reformulated problem is expressed as
\begin{equation} \label{eq:SPMinMulticastApproximatedReform}
\begin{array}{ll}
\displaystyle \underset{\{{\vec W}_{g}\}_{g \in \mathcal{G}}, {\boldsymbol \theta} }{\mathrm{min.}}  &  \displaystyle  \sum\limits_{g \in \mathcal{G}} {\rm Tr} \left({\vec W}_{g}\right)\\
{\mathrm{s.\ t.}}
& \displaystyle \hspace{-0.5cm} \frac{{\rm Tr} \left({\vec H}_{b,u}{\vec W}_{g}\right)}{{\sigma_{u}^{2} + \sum\limits_{j \in \mathcal{B} \setminus \{b\}} \theta_{j,u} + \sum\limits_{k \in \mathcal{G}_{b} \setminus \{g\}} {\rm Tr} \left({\vec H}_{b,u} {\vec W}_{k} \right)}} \geq \gamma_{u}, \\
& \hspace{-0.5cm} \forall b \in \mathcal{B}, \forall g \in \mathcal{G}_b, \forall u \in \mathcal{U}_g \\
& \hspace{-0.5cm} \sum\limits_{i \in \mathcal{G}_{b}} {\rm Tr} \left({\vec H}_{b,u} {\vec W}_{i} \right) \leq \theta_{b,u}, \forall b \in \mathcal{B}, \forall u \in \mathcal{U} \setminus \mathcal{U}_b \\
& \hspace{-0.5cm} {\vec W}_{g} \succeq 0, \forall g \in \mathcal{G}_b, \forall b \in \mathcal{B}
\end{array}
\end{equation}
where $\mathcal{U}_b$ is the user set served by BS $b$, $\theta_{b,u}$ is the ICI from BS $b$ to user $u$, and the vector ${\boldsymbol \theta}$ consists of all ICI variables. Since the inequality constraints are met with equality at the optimal solution, \eqref{eq:SPMinMulticastApproximatedReform} and \eqref{eq:SPMinMulticastApproximated} are equivalent. Problem \eqref{eq:SPMinMulticastApproximatedReform} still requires central processing but it will be the basis for deriving the distributed methods in the next sections.

\subsubsection{Primal Decomposition -based Approach}
Primal decomposition is applicable to decouple an optimization problem which has coupling constraints \cite{Palomar-06}. Specifically, the specific problem decouples by fixing the coupling constraints.
Primal decomposition divides the one-level optimization problem into two levels, i.e., the lower level subproblems and the higher level master problem. Herein, we generalize the method
of [26] to
a multigroup system model.

In the primal decomposition method, \eqref{eq:SPMinMulticastApproximatedReform} is divided into a two-level optimization. Firstly, by fixing the inter-cell interference levels, the beamformers are solved from the BS-specific subproblems by locally minimizing the BS-specific transmit powers. Secondly, the interference levels are optimized by solving a network wide master problem.
The local subproblem for BS $b$ is formulated as an SDP
\begin{equation} \label{eq:SPMinMulticastSubproblem}
\begin{array}{ll}
\displaystyle \underset{\{{\vec W}_{g}\}_{g \in \mathcal{G}_b}}{\mathrm{min.}}  &  \displaystyle  \sum\limits_{g \in \mathcal{G}_b} {\rm Tr} \left({\vec W}_{g}\right)\\
{\mathrm{s.\ t.}}
& \displaystyle \hspace{-0.8cm} \frac{{\rm Tr} \left({\vec H}_{b,u}{\vec W}_{g}\right)}{{\sigma_{u}^{2} + \sum\limits_{j \in \mathcal{B} \setminus \{b\}} \theta_{j,u} + \sum\limits_{k \in \mathcal{G}_{b} \setminus \{g\}} {\rm Tr} \left({\vec H}_{b,u} {\vec W}_{k} \right)}} \geq \gamma_{u}, \\
& \hspace{-0.8cm} \forall g \in \mathcal{G}_b, \forall u \in \mathcal{U}_g \\
& \hspace{-0.8cm} \sum\limits_{i \in \mathcal{G}_{b}} {\rm Tr} \left({\vec H}_{b,u} {\vec W}_{i} \right) \leq \theta_{b,u}, \forall u \in \mathcal{U} \setminus \mathcal{U}_b \\
& \hspace{-0.8cm} {\vec W}_{g} \succeq 0, \forall g \in \mathcal{G}_b,
\end{array}
\end{equation}
which can be optimally solved using any SDP solver.
The network wide master problem can be cast as
\begin{equation} \label{eq:SPMinMulticastMasterProblem}
\begin{array}{cl}  \underset{\{{\boldsymbol \theta}_{b}\}_{b \in \mathcal{B}}}{\mathrm{min.}}  &  \sum\limits_{b \in \mathcal{B}} f^{\star}_{b} ({\boldsymbol \theta}_{b}) \\
{\mathrm{s.\ t.}} & {\boldsymbol \theta}_{b} \in \R^{L}_{++}, \forall b \in \mathcal{B} \\
\end{array}
\end{equation}
where $\R^{L}_{++}$ is a set of positive $L$-dimensional real vectors, and the optimal objective value of \eqref{eq:SPMinMulticastSubproblem} for given ${\boldsymbol \theta}_{b}$ is denoted by $f^{\star}_{b} ({\boldsymbol \theta}_{b})$. The vector ${\boldsymbol \theta}_{b}$ with length $L$ includes the inter-cell interference terms which are associated with BS $b$.
Problem \eqref{eq:SPMinMulticastMasterProblem} to find the inter-cell interference variables $\{\theta_{b,u}\}_{b \in \mathcal{B}, u \in \mathcal{U} \setminus \mathcal{U}_b}$ can be solved by using the projected subgradient method
\begin{eqnarray} \label{eq:SubgradientMethod}
\theta_{b,u}^{(r+1)} & = & \mathcal{P} \left \{ \theta_{b,u}^{(r)} - \varsigma^{(r)} s_{b,u}^{(r)} \right \}, b \in \mathcal{B}, u \in \mathcal{U} \setminus \mathcal{U}_b
\end{eqnarray}
where $\mathcal{P}$ is the projection onto a positive orthant, $r$ is the iteration index, $\varsigma^{(r)}$ is the step-size and $s_{b,u}^{(r)}$ is the subgradient of \eqref{eq:SPMinMulticastMasterProblem} at point $\theta_{b,u}^{(r)}$. Since problem \eqref{eq:SPMinMulticastApproximatedReform} is convex, we can find the subgradient $s_{b,u}^{(r)}$ via the dual problem of \eqref{eq:SPMinMulticastApproximatedReform} as in \cite{Pennanen-14a}. The resulting subgradient at point $\theta_{b,u}^{(r)}$ is given by $s_{b,u}^{(r)} = \lambda_{b,u}^{(r)} - \mu_{j,u}^{(r)}$,
where $\lambda_{b,u}^{(r)}$ is the dual variable associated with $\theta_{b,u}^{(r)}$ in the SINR constraint of user $u$ at its serving BS $b$ (i.e., in subproblem $b$) and $\mu_{j,u}^{(r)}$ is the dual variable associated with $\theta_{b,u}^{(r)}$ in the inter-cell interference constraint of user $u$ at the interfering BS $j$ (i.e., in subproblem $j$).
Because \eqref{eq:SPMinMulticastSubproblem} is an SDP, solving it using standard SDP solvers returns the optimal dual variables as side information (i.e., a certificate for optimality). Another method to find the dual variables is to formulate and solve the dual problem of \eqref{eq:SPMinMulticastSubproblem}.

The step-size of the projected subgradient method has to be properly chosen \cite{Palomar-06} in order to solve the master problem optimally. One example of the valid step size rule is, e.g., a nonsummable diminishing step size, satisfying $\varsigma^{(r)}\geq 0, \lim_{r\to\infty}\varsigma^{(r)}=0$ and $\sum_{r=1}^\infty\varsigma^{(r)}=\infty$.
If all the optimal covariance matrices $\{{\vec W}_{g}^{*}\}_{g \in \mathcal{G}_b, b \in \mathcal{B}}$ returned by the proposed method have unit ranks, then this solution is also optimal for the original problem \eqref{eq:SPMinMulticast}. In this case, the optimal beamformers $\{{\vec w}_{g}^{*}\}_{g \in \mathcal{G}_b, b \in \mathcal{B}}$ are obtained from $\{{\vec W}_{g}^{*}\}_{g \in \mathcal{G}_b, b \in \mathcal{B}}$ by applying the eigenvalue decomposition, i.e., ${\vec w}_{g}^{*}=\sqrt{\lambda_{g}}{\vec u}_{g}$, $\forall g \in \mathcal{G}_b$, $\forall b \in \mathcal{B}$. In case at least one of $\{{\vec W}_{g}^{*}\}_{g \in \mathcal{G}_b, b \in \mathcal{B}}$ has a rank higher than one, then the SDR is not optimal.
Then we can use the Gaussian randomization method presented in Section \ref{sec:GR} to find feasible rank-one beamformers.

\subsubsection*{Distributed Implementation}
\label{sec:SummaryAlg}
When local CSI is available at each BS and the BSs are allowed to exchange scalar information via low-rate backhaul links, we can develop the following distributed beamforming implementation.
The subproblem $b$ in \eqref{eq:SPMinMulticastSubproblem} and the corresponding part of the master problem in \eqref{eq:SPMinMulticastMasterProblem}, i.e., the update of ${\boldsymbol \theta}_{b}$, are solved independently at BS $b$, for all $b \in \mathcal{B}$ in parallel.
At subgradient iteration $r$, BS $b$ signals the dual variables associated with the SINR constraints, i.e., $\{\lambda_{b,u}\}_{u \in \mathcal{U}_{b}}$, to all the interfering BSs using the backhaul links. On the other hand, each BS $b$ signals the dual variables associated with the ICI constraints, i.e., $\{\mu_{b,u}\}_{u \in \mathcal{U} \setminus \mathcal{U}_{b}}$, only to the serving BS of each user $u$, since each ICI variable $\theta_{b,u}$ couples only two BSs, the serving BS of user $u$ and the interfering BS $b$. The total amount of the required backhaul signaling at each subgradient iteration is $2B(B-1)(U/B)$ which is the sum of the real-valued terms exchanged between the coupled BS pairs, assuming a fully connected network and an equal number of users and groups at each cell (i.e., $U_b=U/B$, $\forall b \in \mathcal{B}$).
Although the primal decomposition method results in optimal solution for \eqref{eq:SPMinMulticastApproximatedReform}, the BSs only know their own covariance matrices, while the other BSs may have higher rank covariance matrices. To this end, each BS can send a one-bit feedback to indicate if the rank is higher than 1. If Gaussian randomization is required, the BS-specific powers for each Gaussian randomization instance need to be shared among other BSs to select the best one in a distributed manner.
The overall distributed approach is summarized in {\it Algorithm~\ref{alg:SPMinMulticastAlg}}, which is performed at BS $b$, for all $b$ in parallel. The proposed algorithm applies a subgradient method to update the ICI variables for which the convergence to the optimal solution (for convex problems) has been analyzed, e.g., in \cite{Bertsekas-03}, \cite{Boyd-07}.

\begin{algorithm} [tbp!]
\caption{Primal decomposition based distributed multicast beamforming for power minimization}
\label{alg:SPMinMulticastAlg}
\begin{algorithmic}[1]
\STATE Set $r=0$. Initialize inter-cell interference powers $\boldsymbol{\theta}_{b}^{(0)}$.
\REPEAT
\STATE Compute optimal transmit covariance matrices $\{{\vec W}_{g}^{*}\}_{g \in \mathcal{G}_b}$ and dual variables $\{{\lambda}_{b,u}\}_{u \in \mathcal{U}_b}$, $\{{\mu}_{b,u}\}_{u \in \mathcal{U} \setminus \mathcal{U}_b}$ by solving the relaxed subproblem $b$ as an SDP \eqref{eq:SPMinMulticastSubproblem}.
\STATE Communicate dual variables $\{{\lambda}_{b,u}\}_{u \in \mathcal{U}_b}$, $\{{\mu}_{b,u}\}_{u \in \mathcal{U} \setminus \mathcal{U}_b}$ to the coupled BSs via backhaul.
\STATE Update inter-cell interference variables $\boldsymbol {\theta}_{b}^{(r+1)}$ via projected subgradient method \eqref{eq:SubgradientMethod}.
\STATE Set $r=r+1$.
\UNTIL{desired level of convergence}
\STATE Check whether the ranks of $\{{\vec W}_{g}^{*}\}_{g \in \mathcal{G}_b}$ are all one or not. Share this one-bit information among other BSs via backhaul. If the ranks are one for all $g \in \mathcal{G}_b, b \in \mathcal{B}$, apply eigenvalue decomposition for $\{{\vec W}_{g}^{*}\}_{g \in \mathcal{G}_b}$ to find optimal beamformers $\{{\vec w}_{g}^{*}\}_{g \in \mathcal{G}_b}$ for the original problem. Otherwise, apply Gaussian randomization with power optimization \eqref{eq:GR_PowOpt} to find feasible beamformers $\{{\vec w}_{g}\}_{g \in \mathcal{G}_b}$.
\end{algorithmic}
\end{algorithm}

\subsubsection{ADMM-based Approach}
\label{sec:ADMM}
Next we present the proposed ADMM-based distributed solution to \eqref{eq:SPMinMulticastApproximatedReform}. As a first step, we equivalently reformulate \eqref{eq:SPMinMulticastApproximatedReform} as
\begin{equation} \label{eq:SPMinMulticastLocalCopies}
\begin{array}{ll}
\displaystyle \underset{\{{\vec W}_{g}\}_{g \in \mathcal{G}}, \{\boldsymbol\theta_b,\tilde{\boldsymbol\theta}_b\}_{b\in\mathcal{B}}}{\mathrm{min.}}  &  \displaystyle  \sum\limits_{g \in \mathcal{G}} {\rm Tr} \left({\vec W}_{g}\right)\\
{\mathrm{s.\ t.}}
& \displaystyle \hspace{-2cm} \frac{{\rm Tr} \left({\vec H}_{b,u}{\vec W}_{g}\right)}{{\sigma_{u}^{2} + \sum\limits_{j \in \mathcal{B} \setminus \{b\}} \tilde{\theta}_{j,u}^b + \sum\limits_{k \in \mathcal{G}_{b} \setminus \{g\}} {\rm Tr} \left({\vec H}_{b,u} {\vec W}_{k} \right)}} \geq \gamma_{u}, \\
& \hspace{-2cm} \forall b \in \mathcal{B}, \forall g \in \mathcal{G}_b, \forall u \in \mathcal{U}_g \\
& \hspace{-2cm} \sum\limits_{i \in \mathcal{G}_{b}} {\rm Tr} \left({\vec H}_{b,u} {\vec W}_{i} \right) \leq \tilde{\theta}_{b,u}^b, \forall b \in \mathcal{B}, \forall u \in \mathcal{U} \setminus \mathcal{U}_b \\
& \hspace{-2cm} {\vec W}_{g} \succeq 0, \forall g \in \mathcal{G}_b, \forall b \in \mathcal{B},\\
 &  \hspace{-2cm} \tilde{\boldsymbol\theta}_{b} = \boldsymbol\theta_b, \forall b \in \mathcal{B}
\end{array}
\end{equation}
where we have introduced two new variables $\tilde{\theta}_{b,u}^{b}$ and $\tilde{\theta}_{b,u}^m$ for each ${\theta}_{b,u}$ \cite{Tolli-11,TervoWsumEE-15}. In the first constraint, each BS $b$ has replaced the ICI variable $\theta_{j,u}$ (i.e., the ICI that BS $j$ causes to the user $u$ of BS $b$) with $\tilde{\theta}_{j,u}^b$, i.e., this interference is locally optimized. On the other hand, in the second constraint, each BS $b$ has replaced $\theta_{b,u}$ with $\tilde{\theta}_{b,u}^b$. Now, let us consider a user $u$ which is served by BS $m$. Then, the ICI which BS $b$ causes to user $u$ is $\theta_{b,u}$ but this value is locally optimized by both BS $b$ (this variable is $\tilde{\theta}_{b,u}^b$) and also BS $m$ (this variable is $\tilde{\theta}_{b,u}^m$). The last constraint of (13) then makes sure that $\tilde{\theta}_{b,u}^b=\tilde{\theta}_{b,u}^m$. The introduction of the new variables is required to arrive at a form tractable to the application of ADMM as presented shortly. Intuitively, we can interpret $\tilde{\theta}_{b,u}^{m}$ as the local version of ${\theta}_{b,u}$ optimized by BS $m$. A vector $\tilde{\boldsymbol\theta}_b$ is composed of all the local variables which are associated with BS $b$, i.e., the interference from BS $b$ to the users of other cells and the interference from other BSs $m \in \mathcal{B} \setminus \{b\}$ towards the users of cell $b$. This means that each interference term ${\theta}_{b,u}$ couples only two base stations. Similarly, $\boldsymbol\theta_b$ involves the global variables associated with BS $b$ in a same order as in $\tilde{\boldsymbol\theta}_b$. With the new formulation in \eqref{eq:SPMinMulticastLocalCopies}, let us define
\begin{eqnarray}\label{Localset}
\mathcal{S}_{b} & = & \Bigl\{ (\{{\vec W}_{g}\}_{g \in \mathcal{G}_b}, \tilde{\boldsymbol\theta}_b)\bigr|\nonumber \\
 &  & \frac{{\rm Tr} \left({\vec H}_{b,u}{\vec W}_{g}\right)}{{\sigma_{u}^{2} + \sum\limits_{j \in \mathcal{B} \setminus \{b\}} \tilde{\theta}_{j,u}^b + \sum\limits_{k \in \mathcal{G}_{b} \setminus \{g\}} {\rm Tr} \left({\vec H}_{b,u} {\vec W}_{k} \right)}} \geq \gamma_{u},\nonumber  \\
 &  & \forall g \in \mathcal{G}_b, \forall u \in \mathcal{U}_g\nonumber \\
 &  & \sum\limits_{i \in \mathcal{G}_{b}} {\rm Tr} \left({\vec H}_{b,u} {\vec W}_{i} \right) \leq \tilde{\theta}_{b,u}^b, \forall u \in \mathcal{U} \setminus \mathcal{U}_b\nonumber  \\
 &  & {\vec W}_{g} \succeq 0, \forall g \in \mathcal{G}_b \Bigl\}\end{eqnarray}
to be a local feasible set for cell $b$. As a result, problem \eqref{eq:SPMinMulticastLocalCopies} can be compactly expressed as
\begin{subequations}
\label{eq:EEmax:SCAproblemIntConsensus}
\begin{align}
& \hspace{-0.4cm} \underset{\{{\vec W}_{g}\}_{g \in \mathcal{G}}, \{\boldsymbol\theta_b,\tilde{\boldsymbol\theta}_b\}_{b\in\mathcal{B}}}{\mathrm{min.}} \quad \sum\limits_{g \in \mathcal{G}} {\rm Tr} \left({\vec W}_{g}\right)\label{eq:ADMMobjective}\\
& \hspace{0.5cm} \st  \quad \hspace{1cm}    (\{{\vec W}_{g}\}_{g \in \mathcal{G}_b}, \{\tilde{\boldsymbol\theta}_b\})\in \mathcal{S}_{b}, \forall b \in \mathcal{B}\label{ADMMlocal}\\
& \quad \hspace{2.1cm} \tilde{\boldsymbol\theta}_{b} = \boldsymbol\theta_b, \forall b \in \mathcal{B}\label{ADMMglobal}.
\end{align}
\end{subequations}
At this point, it is observed that \eqref{eq:EEmax:SCAproblemIntConsensus} admits a form of standard global consensus problem which lends itself to the application of ADMM \cite{Boyd-11}. Specifically, the objective function \eqref{eq:ADMMobjective} and the constraints \eqref{ADMMlocal} are separable, and then the consensus constraints \eqref{ADMMglobal} force the equality between the local variables $\tilde{\boldsymbol\theta}_b$ and the global variables $\boldsymbol\theta_b$. Let us define the local variables of BS $b$ as $\boldsymbol\Xi_b = \{(\{{\vec W}_{g}\}_{g \in \mathcal{G}_b}, \{\tilde{\boldsymbol\theta}_b\})\}$, and define
\begin{equation} \label{eq:FuncF}
\begin{array}{ll}
\displaystyle & \hspace{-0.5cm} h_{b}\left(\boldsymbol\Xi_b \right)= \left\{ \begin{array}{ll}
\displaystyle \sum\limits_{g \in \mathcal{G}_b} {\rm Tr} \left({\vec W}_{g}\right) , & \hspace{-0.3cm} \boldsymbol\Xi_b \in \mathcal{S}_{b} \\
\displaystyle \infty, & \hspace{-0.3cm} {\rm otherwise}. \\
 \end{array} \right.
\end{array}
\end{equation}
To apply the ADMM method, we first write
\begin{equation} \label{PartLagr}
\begin{aligned}
& L(\{\boldsymbol\Xi_b,\boldsymbol\theta_b,\boldsymbol\nu_b\}_{\forall b \in \mathcal{B}}) \\ & =  \sum_{b\in\mathcal{B}}\big(h_b(\boldsymbol\Xi_b) + \boldsymbol\nu_{b}\trans(\tilde{\boldsymbol\theta}_b - \boldsymbol\theta_b) + \tfrac{\rho}{2}||\tilde{\boldsymbol\theta}_b - \boldsymbol\theta_b||_2^2\big)
\end{aligned}
\end{equation}
which is a partial augmented Lagrangian of \eqref{eq:EEmax:SCAproblemIntConsensus}. In \eqref{PartLagr}, the dual variables related to the interference equality constraints \eqref{ADMMglobal} are denoted by $\{\boldsymbol\nu_{b}\}_{b\in\mathcal{B}}$. The final term in \eqref{PartLagr} with parameter $\rho > 0$ penalizes for the violation of the equality constraints \eqref{ADMMglobal}.
The variable updates in ADMM can be written as:
\begin{subequations}
\begin{align}
& \text{\emph{Local variable updates:}}\nonumber\\
& \boldsymbol\Xi_b^{(l+1)} = \underset{\boldsymbol\Xi_b}{\argmin} \quad L(\boldsymbol\Xi_b,\boldsymbol\theta_b^{(l)},\boldsymbol{\nu}_b^{(l)}), \forall b \in \mathcal{B} \label{eq:ADMM:primal} \\
& \text{\emph{Global variable updates:}}\nonumber\\
 & \{\boldsymbol\theta_b^{(l+1)}\}_{\forall b\in\mathcal{B}}  = \underset{\{\boldsymbol\theta_b\}_{\forall b\in\mathcal{B}}}{\argmin} \quad L(\{\boldsymbol\Xi_b^{(l+1)},\boldsymbol\theta_b,\boldsymbol{\nu}_b^{(l)}\bigr\}_{\forall b \in \mathcal{B}})\label{eq:ADMM:global} \\
 & \text{\emph{Dual variable updates:}}\nonumber\\
&
\boldsymbol\nu_b^{(l+1)} = \boldsymbol\nu_b^{(l)} + \rho(\tilde{\boldsymbol\theta}_b^{(l+1)} - \boldsymbol\theta_b^{(l+1)}), \forall b \in \mathcal{B}\label{eq:ADMM:dualnu}
\end{align}
\end{subequations}
where $l$ is the iteration index.
From \eqref{eq:ADMM:primal}, the local variables are found from the optimal solution $\boldsymbol\Xi_b^*$ of the following convex problem
\begin{eqnarray}
\label{eq:EEmax:LocalProblem}
\underset{\boldsymbol\Xi_b}{\mathrm{min.}} & h_b(\boldsymbol\Xi_b) + (\boldsymbol\nu_{b}^{(l)})\trans(\tilde{\boldsymbol\theta}_b - \boldsymbol\theta_b^{(l)}) + \tfrac{\rho}{2}||\tilde{\boldsymbol\theta}_b - \boldsymbol\theta_b^{(l)}||_2^2,
\end{eqnarray}
and then the update is written as $\boldsymbol\Xi_b^{(l+1)}=\boldsymbol\Xi_b^*$.
To find the global variables from \eqref{eq:ADMM:global}, the updates are written as $\boldsymbol\theta_b^{(l+1)}=\boldsymbol\theta_b^*$, where $\boldsymbol\theta_b^*$ is the optimal solution of the quadratic unconstrained convex program
\begin{equation}
\underset{\{\boldsymbol\theta_b\}_{b\in\mathcal{B}}}{\mathrm{min.}} \sum_{b\in\mathcal{B}}((\boldsymbol\nu_{b}^{(l)})\trans(\tilde{\boldsymbol\theta}_b^{(l+1)} - \boldsymbol\theta_b) + \tfrac{\rho}{2}||\tilde{\boldsymbol\theta}_b^{(l+1)} - \boldsymbol\theta_b||_2^2).\label{eq:EEmax:GlobalProblem}
\end{equation}
Problem \eqref{eq:EEmax:GlobalProblem} is separable in $\boldsymbol\theta_b$ and can be solved in parallel when the local variables $\tilde{\boldsymbol\theta}_b$ have been exchanged between the coupling BSs. By solving the point of the zero gradient of \eqref{eq:EEmax:GlobalProblem} gives
\begin{eqnarray}
\label{eq:EEmax:ComponentWise}
\theta_{b,j}^* = \frac{1}{2}\big(\tilde{\theta}_{b,j}^{b,(l+1)} + \tilde{\theta}_{b,j}^{m,(l+1)} + \frac{1}{\rho}(\nu_{b,j}^{b,(l)} + \nu_{b,j}^{m,(l)})\big),
\end{eqnarray}
where $m$ and $b$ are the serving and interfering BS of user $j$, respectively. Equation \eqref{eq:EEmax:ComponentWise} can be simplified as $\theta_{b,j}^{(l+1)} = \theta_{b,j}^* = 1/2(\tilde{\theta}_{b,j}^{b,(l+1)} + \tilde{\theta}_{b,j}^{m,(l+1)})$, because inserting $\boldsymbol\theta_b^{(l+1)}$ to \eqref{eq:ADMM:dualnu} results in
\begin{eqnarray}
\nu_{b,j}^{m,(l+1)} = \frac{\rho}{2}(\tilde{\theta}_{b,j}^{m,(l+1)} - \tilde{\theta}_{b,j}^{b,(l+1)}) + \frac{1}{2}(\nu_{b,j}^{m,(l)} - \nu_{b,j}^{b,(l)}) \\
\nu_{b,j}^{b,(l+1)} = \frac{\rho}{2}(\tilde{\theta}_{b,j}^{b,(l+1)} - \tilde{\theta}_{b,j}^{m,(l+1)}) + \frac{1}{2}(\nu_{b,j}^{b,(l)} - \nu_{b,j}^{m,(l)})
\end{eqnarray}
From the above expressions, we can see that $\nu_{b,j}^{m,(l+1)}$ is actually a complement of $\nu_{b,j}^{b,(l+1)}$, i.e., $\nu_{b,j}^{m,(l+1)} + \nu_{b,j}^{b,(l+1)}=0$.
The ADMM-based distributed method is compactly written in Algorithm \ref{algo:ADMMSCA}.
The convergence proof of ADMM to optimal solution, applied to the convex problem of type \eqref{eq:EEmax:SCAproblemIntConsensus}, can be found, e.g., in \cite{Boyd-11}.

\begin{algorithm}[h]
\begin{algorithmic}[1] \caption{ADMM-based distributed multicast beamforming for power minimization.}
\label{algo:ADMMSCA}
\renewcommand{\algorithmicrequire}{\textbf{Initialization:}}
\REQUIRE Set $l=0$, and generate initial points $\boldsymbol\Xi_b^{(0)}$.
\REPEAT \STATE $l:=l+1$.
\STATE Compute optimal transmit covariance matrices $\{{\vec W}_{g}^{*}\}_{g \in \mathcal{G}_b}$ and local copies of inter-cell interference terms $\{\tilde{\theta}_{b,u}^b\}_{u\in\mathcal{U} \setminus \mathcal{U}_b}$ and $\{\tilde{\theta}_{m,j}^b\}_{m\in\mathcal{B}\setminus \{b\}, j\in\mathcal{U}_b}$ using \eqref{eq:ADMM:primal}
\STATE Communicate local copies $\{\tilde{\theta}_{b,u}^b\}_{u\in\mathcal{U} \setminus \mathcal{U}_b}$ and $\{\tilde{\theta}_{m,j}^b\}_{m\in\mathcal{B}\setminus \{b\}, j\in\mathcal{U}_b}$ to the coupled BSs via backhaul.
\STATE Compute global inter-cell interference variables $\{\theta_{b,u}\}_{u\in\mathcal{U} \setminus \mathcal{U}_b}$ and $\{\theta_{m,j}\}_{m\in\mathcal{B} \setminus \{b\}, j\in\mathcal{U}_b}$ from \eqref{eq:ADMM:global}.
\STATE Compute local dual variables $\{\nu_{b,u}\}_{u\in\mathcal{U} \setminus \mathcal{U}_b}$ and $\{\nu_{m,j}\}_{m\in\mathcal{B} \setminus \{b\}, j\in\mathcal{U}_b}$ from \eqref{eq:ADMM:dualnu}.
\UNTIL desired level of convergence
\STATE Step 8 as in Alg. 2.
\end{algorithmic}
\end{algorithm}

\subsubsection*{Distributed Implementation}
As in the primal decomposition based approach, the distributed implementation is enabled if each BS acquires local CSI and scalar information exchange between the BSs is allowed via low-rate backhaul links. The first local subproblems \eqref{eq:ADMM:primal} are completely decoupled and can be solved independently in parallel at each BS, with fixed global variables $\boldsymbol\theta_b$. The update of global variables \eqref{eq:ADMM:global} requires the knowledge of the local copies as shown in \eqref{eq:EEmax:ComponentWise}. At ADMM iteration $l$, BS $b$ signals the local copies $\tilde{\boldsymbol\theta}_b$, to all the interfering BSs. Since each of inter-cell interference term $\theta_{b,u}$ couples only two BSs, the number of exchanged scalars per ADMM iteration is given by $2B(B-1)(U/B)$ which is the same as in the primal decomposition based approach. Finally, each BS can again locally update the dual variables in \eqref{eq:ADMM:dualnu} with the updated $\boldsymbol\theta_b$ from the previous step.
 Similarly to the primal decomposition -based method, a one-bit feedback can be transmitted to indicate the rank of the covariance matrices and possible Gaussian randomization requires the BS-specific powers to be shared between the BSs.
The overall distributed approach is summarized in {\it Algorithm~\ref{algo:ADMMSCA}}, which is performed at BS $b$, for all $b$ in parallel.

\subsubsection{Distributed Gaussian Randomization}
\label{sec:GR}
As mentioned earlier, the solution of the SDR is not optimal to the original problem \eqref{eq:SPMinMulticast} if at least one of $\{{\vec W}_{g}^{*}\}_{g \in \mathcal{G}_b, b \in \mathcal{B}}$ produced by distributed methods has a rank higher than one. In this case, we propose to use the Gaussian randomization method to find feasible beamformers. As in the centralized Gaussian randomization presented in Section \ref{sec:CentralizedAlg}, the optimal covariance matrices returned by the distributed methods are used to generate candidate (normalized) beamforming solutions $\hat{{\vec w}}_{g}$, $\forall g \in \mathcal{G}_b$, $\forall b \in \mathcal{B}$, which are Gaussian random variables with zero mean and covariance ${\vec W}_{g}^{*}$.
To guarantee the feasibility of the beamforming solution, each BS $b$ solves the following LP
\begin{equation} \label{eq:GR_PowOpt}
\begin{array}{ll}
\displaystyle \underset{\{p_{g}\}_{g \in \mathcal{G}_b}}{\mathrm{min.}}  &  \displaystyle  \sum\limits_{g \in \mathcal{G}_b} p_{g}\\
\hspace{-6pt} {\mathrm{s.\ t.}}
& \hspace{-30pt} \displaystyle \frac{p_g \left|{\vec h}_{b,u}\hat{{\vec w}}_{g}\right|^{2}}{{\sigma_{u}^{2} + \sum\limits_{j \in \mathcal{B} \setminus \{b\}} \theta_{j,u} + \sum\limits_{k \in \mathcal{G}_{b} \setminus \{g\}} p_k \left|{\vec h}_{b,u}\hat{{\vec w}}_{k}\right|^{2}}} \geq \gamma_{u}, \\
& \hspace{-30pt} \forall g \in \mathcal{G}_b, \forall u \in \mathcal{U}_g \\
& \hspace{-30pt} \sum\limits_{i \in \mathcal{G}_{b}} p_i \left|{\vec h}_{b,u}\hat{{\vec w}}_{i}\right|^{2} \leq \theta_{b,u}, \forall u \in \mathcal{U} \setminus \mathcal{U}_b. \\
\end{array}
\end{equation}
After solving \eqref{eq:GR_PowOpt}, BS $b$ can define its beamformers by ${\vec w}_{g}=\sqrt{p_{g}^{*}} \hat{{\vec w}}_{g}$, $\forall g \in \mathcal{G}_b$, where $p_{g}^{*}$ is the optimal power associated with the candidate beamformer $\hat{{\vec w}}_{g}$.
The resulting beamformers are sub-optimal for the original problem.
After generating a predefined number of candidate solutions, the one that gives the lowest objective value of the original problem is selected. To find the lowest network-wide objective, the BS-specific power values have to be exchanged between the BSs. Note that the Gaussian randomization can be implemented in a fully distributed manner, since the ICI variables are fixed in \eqref{eq:GR_PowOpt}.
The inter-cell interference values are obtained from the optimal solution of \eqref{eq:SPMinMulticastMasterProblem} and \eqref{eq:EEmax:SCAproblemIntConsensus} for primal decomposition based method and ADMM-based method, respectively.

\subsubsection{Practical Considerations}
\label{sec:PracticalConsiderations}
To acquire optimal performance, {\it Algorithms~\ref{alg:SPMinMulticastAlg} and \ref{algo:ADMMSCA}} need to be run until convergence (provided that the obtained covariance matrices are all rank-one).
However, this is somewhat impractical since the more iterations are run, the higher the signaling/computational load and the longer the caused delay.
In this respect, {\it Algorithms~\ref{alg:SPMinMulticastAlg} and \ref{algo:ADMMSCA}} naturally lend themselves to a practical design where they can be stopped after a limited number of iterations to reduce delay and signaling load. Limiting the number of iterations comes at the cost of increased sum power. In the primal decomposition, since the inter-cell interference levels are fixed at each iteration, feasible beamformers can be computed via the eigenvalue decomposition or the Gaussian randomization procedure, depending on the rank properties of the covariance matrices. In case of ADMM, the local and global interference variables (i.e., $\tilde{\boldsymbol\theta}_b$ and $\boldsymbol\theta_b$) are different during the iterative processing. As a result, the feasibility of the SINR constraints in the original problem \eqref{eq:SPMinMulticastApproximatedReform} is not necessarily guaranteed before the convergence. However, since the global inter-cell interference terms in \eqref{eq:ADMM:global} are inherently feasible, we can always get feasible beamformers even if the local inter-cell interference terms would be infeasible. Specifically, in order to guarantee a feasible beamforming solution at any iteration, each BS can fix the local interference terms $\tilde{\boldsymbol\theta}_b$ to be equal to the global ones $\boldsymbol\theta_b$, and solve \eqref{eq:ADMM:primal} by keeping $\tilde{\boldsymbol\theta}_b$ fixed.

The overhead of backhaul signaling for the centralized and distributed algorithms are compared in Table \ref{tab:SignalingDistributed}.
In the centralized algorithm, we assume that global CSI is made available for each BS by exchanging local CSI of each BS via backhaul links. Thus, the total backhaul signaling load in terms of scalar-valued channel coefficients is given by $2AU(B-1)B$ if we assume equal number of users and groups per cell.
Here, one complex channel coefficient is considered as two real-valued coefficients.
For the distributed algorithm, the total backhaul signaling load is presented per subgradient iteration or ADMM iteration, because the total backhaul signaling per iteration is the same for the two methods.
In Table \ref{tab:SignalingDistributed}, the values inside the brackets denote the percentage of the signaling load that the distributed methods require compared to the centralized algorithm.
It can be observed that significant backhaul signaling savings are achieved with the distributed methods and the savings are significant even when multiple iterations are performed. To give an example about the signaling overhead between the centralized and distributed approaches, let us consider the case where $\{B,U,A\}=\{4,16,16\}$ (i.e., the last row of Table \ref{tab:SignalingDistributed}). In this case, we could run 64 iterations in the distributed methods and the signaling overhead would be still less than in the centralized method. The gap between the methods increases with the network size.
The backhaul signaling overhead can be significantly reduced by limiting the number of iterations. It is worth observing that the signaling overhead of the centralized method scales both with the number of BS antennas $A$ and the number of users $U$, while only with $U$ for the distributed methods. Thus, if the number of antennas is large, the distributed methods require less signaling even when the channels are static for longer time. Also, if the number of antennas grows high, it is obvious that the signaling overhead becomes a problem and it is relevant to reduce the signaling overhead. However, the benefits of the distributed methods become more important in the time-varying channels \cite{Tolli-11} because the BSs can easily acquire local channel information, and they only need to exchange the user-specific scalars. The centralized method would require sharing all the channel information every time when the channels change, causing significant signaling overhead and delays. In practice, building very high rate and low-latency optical fibers for backhauling could enable centralized methods to be practical, but this would result in very high costs for the operators \cite{Tipmongkolsilp-11,Li-17}. However, with smaller backhaul overhead, wider range of capacity-limited backhaul technologies can be applicable, especially when the network dimensions increase \cite{Osseiran:2016:MWC:2988371,Hossain:2011:CCW:2000310,Gesbert-10}.

A distributed approach allows some special case designs where the number of optimization variables is reduced, leading to a lower computational load and even a further decreased signaling overhead. These special case designs come at the cost of somewhat decreased performance. Some of the possible special cases are presented below:
\begin{itemize}
\item Common interference constraint: $\theta_{b,u}=\theta, \forall b \in \mathcal{B}, \forall u \in \mathcal{U} \setminus {\mathcal{U}}_b$.
\item Fixed interference constraints: $\theta_{b,u}=c_{b,u}$, $\forall b \in \mathcal{B}, \forall u \in \mathcal{U} \setminus {\mathcal{U}}_b$, where $c_{b,u}$ is a predefined constant. Does not require any backhaul signaling.
\item Inter-cell interference nulling, i.e., $\theta_{b,u}=0$, $\forall b \in \mathcal{B}, \forall u \in \mathcal{U} \setminus {\mathcal{U}}_b$.
    Does not require any backhaul signaling.
\end{itemize}

\vspace{0.1cm}
\begin{table}[h!]
\centering
\caption{Total backhaul signaling load.}\label{tab:SignalingDistributed}\vspace{2pt}
\vspace{-0.3cm}
\begin{tabular}{c|c|c|c}
& Centralized & \pbox{20cm}{Distributed \\ (1 iter)} & \pbox{20cm}{Distributed \\ (10 iter)} \\ \hline
$\{B,U,A\}=\{2,8,8\}$ & 256 & 16 (6.3$\%$) & 160 (63$\%$) \\ \hline
$\{B,U,A\}=\{3,12,12\}$ & 1728 & 48 (2.8$\%$) & 480 (28$\%$) \\ \hline
$\{B,U,A\}=\{4,16,16\}$ & 6144 & 96 (1.6$\%$) & 960 (16$\%$) \\ \hline
\end{tabular}
\end{table}

\section{SINR Balancing Problem}
\label{sec:SINRbalancing} 
\subsection{Centralized Beamforming Design} 
Problem \eqref{eq:SINRbalancingMulticast0} is a nonconvex max-min fractional program and difficult to solve as such. To find a more tractable formulation, we first write the equivalent transformation of \eqref{eq:SINRbalancingMulticast0} as
\begin{equation} \label{eq:SINRbalancingMulticast2}
\begin{aligned}
\hspace{-20pt} \displaystyle \underset{t, \{{\vec w}_{g}\}_{g \in \mathcal{G}}}{\mathrm{max.}} & \ \ \displaystyle  t\\
{\mathrm{s.\ t.}}
& \ \  \displaystyle t \leq \frac{|{\vec h}_{b,u}\herm {\vec w}_{g}|^{2}}{{\sigma_{u}^{2} + \sum\limits_{j \in \mathcal{B}} \sum\limits_{k \in \mathcal{G}_j \setminus \{g\}} |{\vec h}_{j,u}\herm {\vec w}_{k}|^{2}}}, \\
& \ \ \forall b \in \mathcal{B}, \forall g \in \mathcal{G}_b, \forall u \in \mathcal{U}_g \\
& \ \ \sum\limits_{g\in\mathcal{G}_b}{\rm Tr}(\mathbf{w}_g\mathbf{w}_g\herm) \leq P_b, \forall b \in \mathcal{B}.
\end{aligned}
\end{equation}
Problems \eqref{eq:SINRbalancingMulticast0} and \eqref{eq:SINRbalancingMulticast2} are equivalent since the first constraint of \eqref{eq:SINRbalancingMulticast2} holds with equality at optimum.
By applying the SDR, we get
\begin{eqnarray}\label{eq:SINRbalancingMulticast3}
&  & \hspace{-16pt}  \underset{t, \{{\vec W}_{g}\}_{g \in \mathcal{G}}}{\mathrm{max.}}   \ \   t \nonumber\\
\hspace{-10pt} {\mathrm{s.\ t.}}
&  & \ \ \displaystyle t \leq \frac{{\rm Tr} \left({\vec H}_{b,u}{\vec W}_{g}\right)}{{\sigma_{u}^{2} + \sum\limits_{j \in \mathcal{B}} \sum\limits_{k \in \mathcal{G}_j \setminus \{g\}} {\rm Tr} \left({\vec H}_{j,u} {\vec W}_{k} \right)}}, \nonumber\\
&  & \ \ \forall b \in \mathcal{B}, \forall g \in \mathcal{G}_b, \forall u \in \mathcal{U}_g\nonumber \\
&  & \ \ \sum\limits_{g\in\mathcal{G}_b}{\rm Tr}(\mathbf{W}_g) \leq P_b, \forall b \in \mathcal{B} \nonumber \\
&  & \ \ {\vec W}_{g} \succeq 0, \forall g \in \mathcal{G}.\nonumber\\
\end{eqnarray}
The above problem is an epigraph form of a quasiconvex optimization problem which can be optimally solved by a bisection search over $t$ \cite{Boyd-04}. Specifically, in the beginning, we initialize lower bound $\underline{t}$ and upper bound $\bar{t}$ and set $t=1/2(\underline{t}+\bar{t})$. For fixed $t$, we solve problem
\begin{equation}
\begin{aligned}\label{eq:SINRbalancingMulticastFeas}
\hspace{-14pt} \displaystyle {\mathrm{find}}  & \ \ \displaystyle \hspace{2pt} \{{\vec W}_{g}\}_{g \in \mathcal{G}}\\
\hspace{-14pt} {\mathrm{s.\ t.}}
&  \ \ \displaystyle t \leq \frac{{\rm Tr} \left({\vec H}_{b,u}{\vec W}_{g}\right)}{{\sigma_{u}^{2} + \sum\limits_{j \in \mathcal{B}} \sum\limits_{k \in \mathcal{G}_j \setminus \{g\}} {\rm Tr} \left({\vec H}_{j,u} {\vec W}_{k} \right)}},\\
& \ \ \forall b \in \mathcal{B}, \forall g \in \mathcal{G}_b, \forall u \in \mathcal{U}_g  \\
& \ \ \sum\limits_{g\in\mathcal{G}_b}{\rm Tr}(\mathbf{W}_g) \leq P_b, \forall b \in \mathcal{B}  \\
& \ \ {\vec W}_{g} \succeq 0, \forall g \in \mathcal{G}
\end{aligned}
\end{equation}
which is an SDP feasibility problem. If the problem is feasible, we set $\underline{t}=t$. Otherwise, we set $\bar{t}=t$. Then we calculate new $t$ as $t=1/2(\underline{t}+\bar{t})$ and repeat until $\bar{t}-\underline{t}\leq\epsilon$, where $\epsilon$ is a small threshold. It is worth mentioning that the same optimality/rank analysis as in power minimization problem applies also to the SINR balancing problem so that if the solution is not rank 1, we can use Gaussian randomization to find a suboptimal solution. In this case, instead of solving \eqref{eq:GR_PowOpt_Centr} as in the power minimization problem, we solve the quasi-linear program
\begin{equation}
\begin{aligned}\label{eq:SINRbalancingGR}
\hspace{-14pt} \displaystyle \underset{t, \{p_{g}\}_{g \in \mathcal{G}}}{\mathrm{max.}}  & \ \ \displaystyle  \hspace{2pt} t \\
\hspace{-8pt} {\mathrm{s.\ t.}}
& \ \ \displaystyle t \leq \frac{p_g|\mathbf{h}_{b,u}\hat{\mathbf{w}}_{g}|^2}{\sigma_{u}^{2} + \sum\limits_{j \in \mathcal{B}} \sum\limits_{k \in \mathcal{G}_j \setminus \{g\}} p_k|\mathbf{h}_{j,u} \hat{\mathbf{w}}_{k}|^2 }, \\
& \ \ \forall b \in \mathcal{B}, \forall g \in \mathcal{G}_b, \forall u \in \mathcal{U}_g  \\
& \ \ \sum\limits_{g\in\mathcal{G}_b}p_g \leq P_b, \forall b \in \mathcal{B}
\end{aligned}
\end{equation}
for a given set of normalized candidate beamformers $\{\hat{\mathbf{w}}_g\}_{g\in\mathcal{G}}$ and choose the one that gives the maximum $t$. Problem \eqref{eq:SINRbalancingGR} can again be solved via the bisection method. The centralized algorithm is summarized in Algorithm \ref{alg:SINRMulticastAlgCentr}. \vspace{-2mm}

\begin{algorithm} [h!]
\caption{Centralized multicast SINR balancing}
\label{alg:SINRMulticastAlgCentr}
\begin{algorithmic}[1]
\STATE Initialize $\underline{t}$ and $\bar{t}$
\REPEAT
\STATE Set $t=1/2(\underline{t}+\bar{t})$
\STATE Compute SDP feasibility problem \eqref{eq:SINRbalancingMulticastFeas} with fixed $t$.
\STATE If \eqref{eq:SINRbalancingMulticastFeas} is feasible, set $\underline{t}=t$, else, set $\bar{t}=t$
\UNTIL $\bar{t}-\underline{t}< \epsilon$
\STATE Check the ranks of $\{{\vec W}_{g}^{*}\}_{g \in \mathcal{G}}$. If the ranks are all one, apply eigenvalue decomposition for $\{{\vec W}_{g}^{*}\}_{g \in \mathcal{G}}$ to find optimal beamformers $\{{\vec w}_{g}^{*}\}_{g \in \mathcal{G}}$ for the original problem. Otherwise, apply Gaussian randomization using \eqref{eq:SINRbalancingGR} to find feasible beamformers $\{{\vec w}_{g}\}_{g \in \mathcal{G}}$.
\end{algorithmic}
\end{algorithm}

\subsection{Distributed Beamforming Design} 

Finding a distributed solution for the SINR balancing problem is more challenging because in addition to the inter-cell interference terms as in the power minimization problem, global variable $t$ couples all the cells. Here we propose a simple distributed design where each cell maximizes its own minimum SINR with the constraint that the interference towards all the other cells are kept below a certain threshold. This method can be realized without any backhaul information exchange. Specifically, each cell locally solves the problem
\begin{eqnarray}\label{eq:SINRbalanceMulticastDistributed}
& & \underset{t_b, \{{\vec W}_{g}\}_{g \in \mathcal{G}_b}}{\mathrm{max.}}   \ \  t_b\nonumber \\
& & \ \  {\mathrm{s.\ t.}}
\  \ \displaystyle t_b \leq \frac{{\rm Tr} \left({\vec H}_{b,u}{\vec W}_{g}\right)}{{\sigma_{u}^{2} + \sum\limits_{j \in \mathcal{B} \setminus \{b\}} \theta_{j,u} + \sum\limits_{k \in \mathcal{G}_{b} \setminus \{g\}} {\rm Tr} \left({\vec H}_{b,u} {\vec W}_{k} \right)}}, \nonumber\\
& & \forall g \in \mathcal{G}_b, \forall u \in \mathcal{U}_g \nonumber\\
& & \sum_{g\in\mathcal{G}_b} {\rm Tr}(\mathbf{H}_{b,u}\mathbf{W}_g) \leq \theta_{b,u}, \forall u \in \mathcal{U} \setminus \mathcal{U}_b \nonumber \\
& & \sum_{g\in\mathcal{G}_b}{\rm Tr}(\mathbf{W}_g) \leq P_b, {\vec W}_{g} \succeq 0, \forall g \in \mathcal{G}_b\nonumber\\
\end{eqnarray}
where $\theta_{j,u}$ is the maximum predefined inter-cell interference from BS $j$ to user $u$. The second constraint makes sure that the interference is kept below the limit $\theta_{j,u}$. In this method, each cell gets slightly different values for the minimum SINR but it is simple and can achieve good performance if the interference levels are chosen reasonably. In practice, these values could be averaged over longer time
and we could then easily use some average values. Testing e.g., long-term values for
different system parameters in offline and using look-up tables to pick up good average
values could be a practical option. In the distributed approach, the Gaussian randomization can be performed locally because the maximum inter-cell interference levels are fixed. Specifically, each BS uses the bisection method to solve the problem
\begin{equation} \label{eq:GR_PowOpt_SINR}
\begin{array}{ll}
& \hspace{-10pt} \displaystyle \underset{t_b, \{p_{g}\}_{g \in \mathcal{G}_b}}{\mathrm{max.}}  \ \  \displaystyle  t_b \\
& \hspace{-18pt} {\mathrm{s.\ t.}}
\ \ \displaystyle t_b \leq \frac{p_g \left|{\vec h}_{b,u}\hat{{\vec w}}_{g}\right|^{2}}{{\sigma_{u}^{2} + \sum\limits_{j \in \mathcal{B} \setminus \{b\}} \theta_{j,u} + \sum\limits_{k \in \mathcal{G}_{b} \setminus \{g\}} p_k \left|{\vec h}_{b,u}\hat{{\vec w}}_{k}\right|^{2}}}, \\
& \hspace{8pt}  \forall g \in \mathcal{G}_b, \forall u \in \mathcal{U}_g \\
& \hspace{8pt}  \sum\limits_{i \in \mathcal{G}_{b}} p_i \left|{\vec h}_{b,u}\hat{{\vec w}}_{i}\right|^{2} \leq \theta_{b,u}, \forall u \in \mathcal{U} \setminus \mathcal{U}_b\\
& \hspace{8pt}  \sum_{g\in\mathcal{G}_b}p_g \leq P_b
\end{array}
\end{equation}
for a given set of normalized candidate beamformers $\{\hat{\mathbf{w}}_g\}_{g\in\mathcal{G}}$ and the one that gives the maximum $t_b$ is chosen at each BS $b$.
The distributed algorithm is summarized in Algorithm \ref{alg:SINRMulticastAlgDecentr}.

\begin{algorithm} [h!]
\caption{Distributed multicast SINR balancing}
\label{alg:SINRMulticastAlgDecentr}
\begin{algorithmic}[1]
\STATE Fix $\theta_{b,u}, \forall b \in \mathcal{B}, u \in \mathcal{U} \setminus \mathcal{U}_b$
\STATE $\forall b \in \mathcal{B}$: Initialize $\underline{t}_b$ and $\bar{t}_b$
\STATE $\textbf{repeat}$ $\forall b \in \mathcal{B}$ in parallel
\STATE \hspace{6pt}  Set $t_b=1/2(\underline{t}_b+\bar{t}_b)$
\STATE \hspace{6pt}  Compute local problem \eqref{eq:SINRbalanceMulticastDistributed} with fixed $t_b$.
\STATE  \hspace{6pt}   If \eqref{eq:SINRbalanceMulticastDistributed} with fixed $t_b$ is feasible, set $\underline{t}_b=t$, else, set $\bar{t}_b=t$
\STATE $\textbf{until}$ $\bar{t}_b-\underline{t}_b< \epsilon$
\STATE Check the ranks of $\{{\vec W}_{g}^{*}\}_{g \in \mathcal{G}}$. If the ranks are all one, apply eigenvalue decomposition. Otherwise, apply local Gaussian randomization using \eqref{eq:GR_PowOpt_SINR} to find feasible beamformers $\{{\vec w}_{g}\}_{g \in \mathcal{G}}$.
\end{algorithmic}
\end{algorithm}

\section{Simulation Results}
\label{sec:SimulationResults}
This section numerically evaluates the performances of the proposed algorithms.
In the simulation model, we assume a network of $B$ BSs with $A$ transmit antennas per each BS. The total number of groups in the network is denoted by $G$, and the groups are then equally divided between the BSs. The number of users per each group is given by $U/G$, where $U$ is the total number of users in the network. The channel conditions are assumed to be frequency-flat Rayleigh fading, and the channel coefficients between the antennas are uncorrelated. We define cell separation parameter $d$ to represent the average path loss between BS $b$ and users of neighboring cells of BS $b$. This means that the interference power towards the users of other cells is attenuated by the value of $d$. More specifically, the observed channel from BS $b$ to user $u$ ($u\in\mathcal{U}\setminus \mathcal{U}_b$) is $\tilde{\mathbf{h}}_{b,u}\triangleq \sqrt{d^{-1}}\mathbf{h}_{b,u}$, where each element of $\mathbf{h}_{b,u}$ is
an i.i.d. complex Gaussian random variable with zero mean and unit variance.  If $d=1$ ($d=0$ dB), the system models the case where all the users are on the cell edges, since the inter-cell interference signals (on average) are as strong as the desired signals.
In all the figures, the main system parameters are given by $\{B,G,U,A\}$.
The SINR constraints are set equal for all users, i.e., $\gamma_u=\gamma$, $\forall u \in \mathcal{U}$. The step size parameter in Algorithm 2 is fixed to $\varsigma^{(r)}=\varsigma=0.3$ and the penalty parameter in Algorithm 3 to $\rho=2$. For Algorithm \ref{alg:SINRMulticastAlgDecentr}, we set the maximum inter-cell interference level to be the same towards all the users, i.e., $\theta_{b,j}=\theta, \forall b \in \mathcal{B}, u \in \mathcal{U} \setminus \mathcal{U}_g$.  In the simulations, we have generated $100$ Gaussian randomizations if the solution has been higher thank rank-1.
First, the convergence behavior of the distributed algorithms for the power minimization are examined, and their performances after limited number of iterations are compared to the centralized approach.
Then, the superiority of coordinated multicast beamforming (i.e., the proposed centralized algorithm for the power minimization) over conventional transmission schemes is demonstrated.
We also compare the performances of the proposed methods to the lower bound solution under different system settings. Furthermore, the tightness of the SDR method and the properties of the higher rank solutions are also examined. Finally, the performances of the proposed methods for the SINR balancing problem are studied with different parameters.

\begin{figure}[tbp!]
  \centering
  \includegraphics[width=0.68\columnwidth]{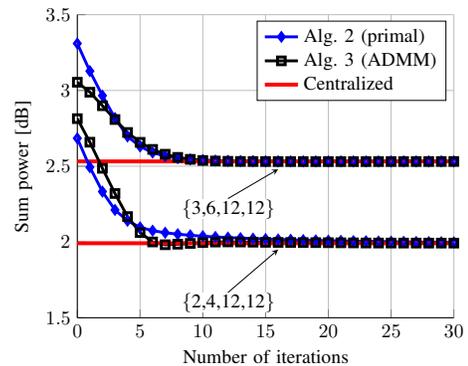}
 \caption{Convergence behavior of distributed algorithms with $\gamma=-1$ dB, $d=1$ dB.}
\label{fig:Convergence}
\end{figure}

\begin{figure}[tbp!]
\centering
\subfigure[Performance after 1 iteration.]{\label{fig:Centr_vs_Distr_a}\includegraphics[width=0.68\columnwidth]{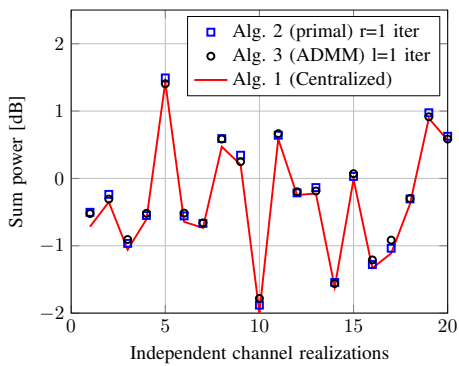}}
\subfigure[Performance after 10 iterations.]{\label{fig:Centr_vs_Distr_b}\includegraphics[width=0.68\columnwidth]{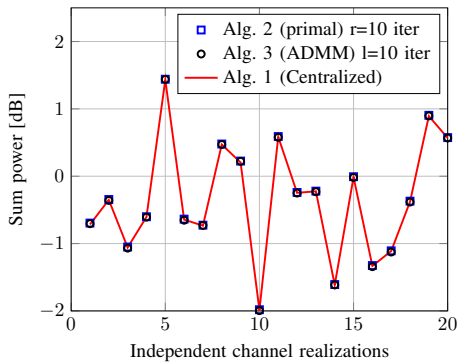}}
\caption{The performances of the distributed algorithms with limited number of iterations with \{2,4,8,12\}, $d=1$ dB, $\gamma=1$ dB.}
\label{fig:Centr_vs_Distr}
\end{figure}

In Fig.~\ref{fig:Convergence}, the convergence behavior of the distributed algorithms is examined under different system settings.
In this example, the speed of convergence is relatively fast. Especially, the first few iterations improve the performance significantly, and after $10$ iterations both algorithms have almost converged. Note that the ADMM-based method can be infeasible for intermediate iterations, as discussed in Section \ref{sec:ADMM}.

In Fig.~\ref{fig:Centr_vs_Distr_a} and ~\ref{fig:Centr_vs_Distr_b}, we consider 20 independent channel realizations and illustrate the performance of the distributed algorithms when the number of iterations is limited to 1 and 10, respectively. The main system parameters are given by $\{B,G,U,A\}=\{2,4,8,12\}$. The results demonstrate that the performances of the distributed algorithms after 10 iterations are very close to that of the centralized scheme. It can be seen that the performances are relatively good even after 1 iteration. We can observe that the better method after 1 iteration depends on the channel realization. It is worth mentioning that the results depend on the initial points and step size values. All the covariance matrices in these results were rank-one.

\begin{figure}[tbp!]
  \centering
  \includegraphics[width=0.68\columnwidth]{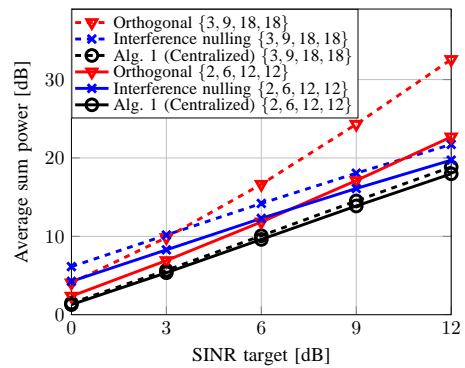}
 \caption{Average sum power versus SINR target for different transmission schemes with $d=1$ dB.}
\label{fig:PvsSINR}
\end{figure}

Fig.~\ref{fig:PvsSINR} plots the average sum power against SINR target for various transmission schemes under different system settings.
The following schemes are compared:
\begin{itemize}
\item Single-cell beamforming with orthogonal access (extension of unicast case in \cite{Tolli-09c} to multicast)
\item Coordinated beamforming with inter-cell interference nulling (proposed special case design in Section \ref{sec:PracticalConsiderations})
\item Coordinated beamforming with inter-cell interference optimization (proposed Algorithm \ref{alg:SPMinMulticastAlgCentr} in Section \ref{sec:CentralizedAlg})
\end{itemize}
In the orthogonal access scheme, each BS uses independent time or frequency slot to optimize the beamformers for its own users, yielding an inter-cell interference free communication scenario.
However, in order to guarantee the same rate targets as in the proposed scheme, the user specific rate targets have to be $B$ times higher than in the non-orthogonal multi-cell case. In the inter-cell interference nulling scheme, each BS forces the inter-cell interference to zero.
For simplicity, the results for coordinated beamforming in Fig.~\ref{fig:PvsSINR} were obtained via centralized processing.
However, the same results can be achieved via distributed algorithms if they are let to converge.
The numerical results illustrate the superiority of the proposed coordinated beamforming method over the conventional transmission schemes.
Significant performance gains over the interference nulling scheme are witnessed especially for low and medium SINR targets. The gain diminishes with the increasing SINR target. On the other hand, the suboptimality of the orthogonal access scheme is greatly emphasized as the SINR target or the number of BSs increases.

\begin{figure}[tbp!]
  \centering
  \includegraphics[width=0.68\columnwidth]{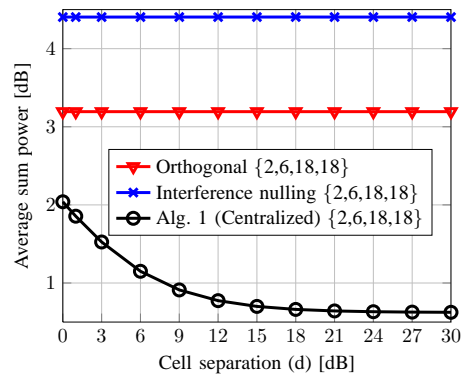}
 \caption{Average sum power versus cell separation for different transmission schemes with $\gamma=1$ dB.}
\label{fig:Pvsd}
\end{figure}

Fig.~\ref{fig:Pvsd} demonstrates the sum power versus cell separation for different transmission schemes. It is observed that the gain of the proposed method increases with cell separation. When cell separation increases, the cells become more isolated and the effect of wasting the degrees of freedom in interference nulling and orthogonal method becomes even more significant.

\begin{figure}[tbp!]
  \centering
  \includegraphics[width=0.68\columnwidth]{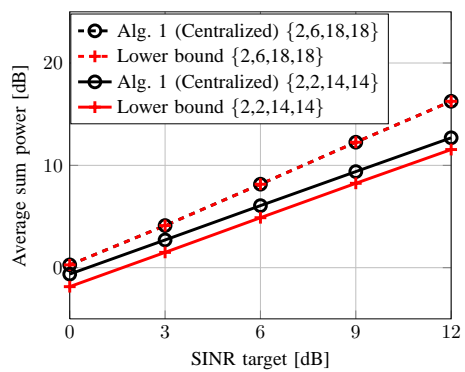}
 \caption{Average sum power versus SINR target with $d=3$ dB.}
\label{fig:PvsSINR2}
\end{figure}

\begin{figure}[tbp!]
  \centering
  \includegraphics[width=0.68\columnwidth]{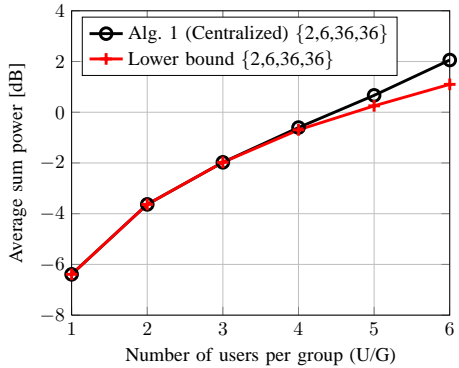}
 \caption{Average sum power versus the number of users per group with $\gamma=1$ dB, $d=1$ dB.}
\label{fig:PvsUsersPerGroup}
\end{figure}

In Fig.~\ref{fig:PvsSINR2}, the centralized algorithm is compared to the lower bound solution of the relaxed problem. Let us remind that in case of rank-one covariance matrices, the SDR is optimal and results in the lower bound. Otherwise, the Gaussian randomization is applied to get a feasible rank-one solution.
The results imply that if the number of users per group is low, (i.e., $U/G=3$), the SDR (usually) finds an optimal solution irrespective of the SINR target. For high number of users per group (i.e., $U/G=7$), some solutions of the SDR problem have rank higher than one.
Hence, the Gaussian randomization yields a small gap between the feasible rank-one result and the lower bound solution.
In Fig.~\ref{fig:PvsUsersPerGroup}, the effect of increasing the number of users per group is further studied. Specifically, sum power is presented against the number of users per group.
For low number of users, it seems that the SDR is optimal since it gives the same solution as the lower bound.
However, when $U/G>4$, there is a minor performance gap between the solutions in the considered setting.

Table \ref{tab:Rank1} presents the probability of rank-one solution in \eqref{eq:SPMinMulticastApproximated} for various number of users per group and for different SINR target values with $d=1$ dB. The results were obtained by averaging over $5000$ channel realizations using simulation parameters $\{B,G,U,A\}=\{2,4,4-24,24\}$.
One can see that the probability decreases with the increasing number of users per group, while the SINR target has less impact. For example, the probability is $100 \%$ for $U/G=1$ and $U/G=2$, while it is less than $25 \%$ for $U/G=6$.
Table \ref{tab:AveRank} further illustrates the average ranks of the higher rank solutions in \eqref{eq:SPMinMulticastApproximated} using the same system parameters as in Table \ref{tab:Rank1}. The average rank is calculated by summing the ranks of all transmit covariance matrices and dividing it by the number of groups $G$, and then averaging it over $5000$ channel realizations. It can be seen that the average rank slightly increases as the number of users per group increases. Since the dimension of each transmit covariance matrix is 24, the maximum rank could be 24. However, the results demonstrate that the average ranks are relatively low, i.e., always below $1.36$. 


\begin{table}[h!]
\centering
\caption{Probability of rank-one solutions ($\%$).}\label{tab:Rank1}
\begin{tabular}{c|c|c|c|c|c|c}
$U/G$ & 1 & 2 & 3 & 4 & 5 & 6 \\ \hline
$\gamma=1$ dB   & 100 & 100 & 97.72 & 74.34 & 45.62 & 22.20 \\ \hline
$\gamma=5$ dB   & 100 & 100 & 97.08 & 76.34 & 46.32 & 23.86 \\ \hline
$\gamma=15$ dB  & 100 & 100 & 98.42 & 76.96 & 45.18 & 21.92 \\ \hline
$\gamma=25$ dB  & 100 & 100 & 99.50 & 74.84 & 40.14 & 16.48 \\ \hline
\end{tabular}
\end{table}

\begin{table}[h!]
\centering
\caption{Average rank of higher rank solutions.}\label{tab:AveRank}
\begin{tabular}{c|c|c|c|c|c|c}
$U/G$ & 1 & 2 & 3 & 4 & 5 & 6 \\ \hline
$\gamma=1$ dB   & - & - & 1.0057 & 1.0703 & 1.1761 & 1.3063 \\ \hline
$\gamma=5$ dB   & - & - & 1.0073 & 1.0650 & 1.1725 & 1.2966 \\ \hline
$\gamma=15$ dB  & - & - & 1.0039 &  1.0619 & 1.1770 & 1.3099 \\ \hline
$\gamma=25$ dB  & - & - & 1.0013 & 1.0692 & 1.2029 & 1.3573 \\ \hline
\end{tabular}
\end{table}

\begin{figure}[tbp!]
  \centering
  \includegraphics[width=0.68\columnwidth]{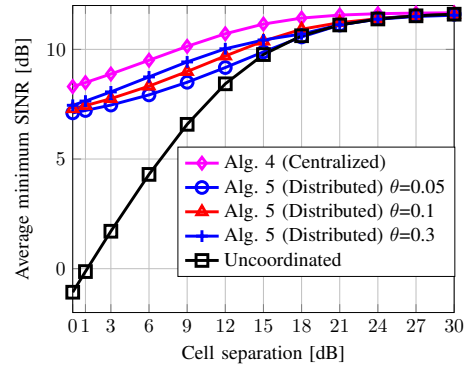}
 \caption{Average minimum SINR versus cell separation with \{2,6,12,12\} and $P_b=10$ W.}
\label{fig:MinimumSINR}
\end{figure}

Fig.~\ref{fig:MinimumSINR} compares the performance of the centralized and distributed algorithms for the SINR balancing problem by plotting the minimum SINR of the users versus cell separation. We compare the performance of the centralized and distributed method by using different interference levels $\theta$. As can be observed, the performance depends on the maximum allowed interference level and cell separation. We can see that the proposed algorithms offer major performance gains over the uncoordinated method when the cell separation is 0-15 dB. We can also see that the distributed schemes achieve only slightly worse performance than the centralized method. When the cell separation is high, all the methods perform equally. This is because with high cell separation, the cells become isolated and there is no need to account for the inter-cell interference.

\begin{figure}[tbp!]
  \centering
  \includegraphics[width=0.68\columnwidth]{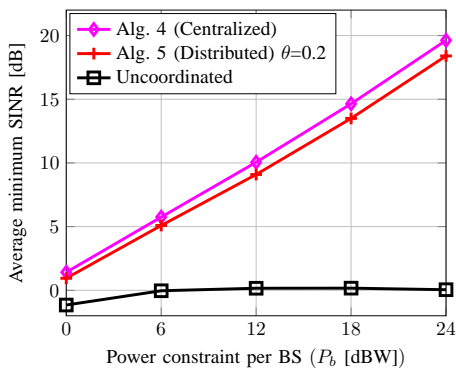}
 \caption{Average minimum SINR versus BS-specific power constraint with \{2,6,12,12\} and $d=1$ dB.}
\label{fig:SINRvsPC}
\end{figure}

\vspace{-4.3mm}
Fig.~\ref{fig:SINRvsPC} plots the minimum SINR versus power constraint for different methods. We compare the proposed methods with the uncoordinated method, where the BSs do not coordinate and do not account for the inter-cell interference in the beamforming optimization. More specifically, each BS solves the following problem:
\begin{equation}
\begin{aligned}\label{eq:UnCoordSINR}
& \hspace{0pt} \displaystyle \underset{t_b, \{{\vec W}_{g}\}_{g \in \mathcal{G}_b}}{\mathrm{max.}}  \ \  \displaystyle  t_b \\
& {\mathrm{s.\ t.}}
 \ \ \displaystyle t_b \leq \frac{{\rm Tr} \left({\vec H}_{b,u}{\vec W}_{g}\right)}{{\sigma_{u}^{2} + \sum\limits_{k \in \mathcal{G}_b \setminus \{g\}} {\rm Tr} \left({\vec H}_{b,u} {\vec W}_{k} \right)}}, \forall g \in \mathcal{G}_b, \forall u \in \mathcal{U}_g \\
& \hspace{20pt} \sum_{g\in\mathcal{G}_b}{\rm Tr}(\mathbf{W}_g) \leq P_b, {\vec W}_{g} \succeq 0, \forall g \in \mathcal{G}_b 
\end{aligned}
\end{equation}
We can see that the performances of all the proposed methods increase with power constraint. Also, the gap between the proposed methods and uncoordinated method increase with larger power constraint. In fact, the performance of the uncoordinated method starts to decrease with larger $P_b$. This is mainly because the uncoordinated method uses more power per BS without taking into account the inter-cell interference which means that the inter-cell interference increases.

\begin{figure}[tbp!]
  \centering
  \includegraphics[width=0.68\columnwidth]{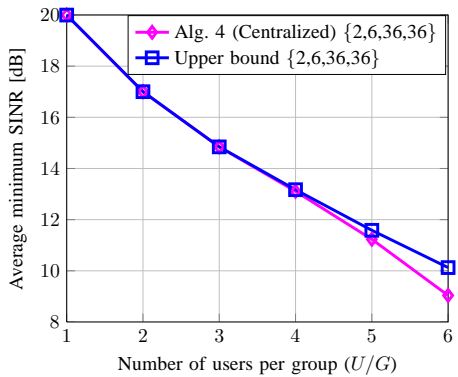}
 \caption{Average minimum SINR versus the number of users per group with $P_b=$10 W, $d=1$ dB.}
\label{fig:SINRvsUpBound}
\end{figure}

Finally, Fig.~\ref{fig:SINRvsUpBound} illustrates the effect of the number of users per group on the rank-1 solution of SINR balancing problem. Especially, here we show the average minimum SINR versus number of users per group. The proposed algorithm is compared to the upper bound solution which is the one produced by the SDR relaxation. When the solution of SDR is rank-1, the two methods results in the same optimal value. In case of higher rank solution, the proposed algorithm performs Gaussian randomization to find a feasible but suboptimal solution. Similarly to Fig. \ref{fig:PvsUsersPerGroup}, the SDR is optimal to the original problem when the number of users per group is low ($U/G\leq4$). However, with $U/G>4$, there is a small performance gap between the methods.

\section{Conclusions}
\label{sec:Conclusion}
This paper has considered multicast beamforming in a multi-cell system, where each multiantenna BS serves multiple groups of single-antenna users. Each distinct group is served by a single beam with common data to all the users in the group. Two optimization goals have been considered. The first one was to minimize the sum transmission power subject to the per-user minimum SINR constraints. A convex approximation of this non-convex problem was obtained via the standard SDR method. The resulting SDP can be solved in a centralized manner if global CSI is available. After presenting the centralized solution, two alternative distributed algorithms were proposed, one of which was based on the primal decomposition and the other on the alternating direction method of multipliers. The implementation of the distributed algorithms requires only local CSI at each BS and low-rate backhaul information exchange between BSs. Coordinated beamforming design was further extended to SINR balancing problem by proposing a centralized and a simple distributed algorithm. These algorithms exploited a bisection method, in which an SDP feasibility problem needs to be solved at each iteration. Beamforming coordination has been shown to be beneficial in comparison to conventional non-coordinated transmissions as demonstrated through numerical examples. The proposed distributed algorithms for power minimization were also shown to achieve nearly the same performance as the centralized approach even after few iterations. We have also illustrated that the SDR provides optimal solutions when the number of users per group is not too high.




\bibliographystyle{IEEEbib}

\end{document}

%% file: commands.tex
\newcommand{\be}{\begin{equation}}
\newcommand{\ee}{\end{equation}}
\newcommand{\ba}{\begin{array}}
\newcommand{\ea}{\end{array}}
\newcommand{\bea}{\begin{eqnarray}}
\newcommand{\eea}{\end{eqnarray}}

\newcommand{\herm}{^{\mbox{\scriptsize H}}}

\newcommand{\vbar}{\raisebox{.17ex}{\rule{.04em}{1.35ex}}}
\newcommand{\vbarind}{\raisebox{.01ex}{\rule{.04em}{1.1ex}}}
\newcommand{\R}{\ifmmode {\rm I}\hspace{-.2em}{\rm R} \else ${\rm I}\hspace{-.2em}{\rm R}$ \fi}
\newcommand{\T}{\ifmmode {\rm I}\hspace{-.2em}{\rm T} \else ${\rm I}\hspace{-.2em}{\rm T}$ \fi}
\newcommand{\N}{\ifmmode {\rm I}\hspace{-.2em}{\rm N} \else \mbox{${\rm I}\hspace{-.2em}{\rm N}$} \fi}
\newcommand{\B}{\ifmmode {\rm I}\hspace{-.2em}{\rm B} \else \mbox{${\rm I}\hspace{-.2em}{\rm B}$} \fi}
\newcommand{\Hil}{\ifmmode {\rm I}\hspace{-.2em}{\rm H} \else \mbox{${\rm I}\hspace{-.2em}{\rm H}$} \fi}
\newcommand{\C}{\ifmmode \hspace{.2em}\vbar\hspace{-.31em}{\rm C} \else \mbox{$\hspace{.2em}\vbar\hspace{-.31em}{\rm C}$} \fi}
\newcommand{\Cind}{\ifmmode \hspace{.2em}\vbarind\hspace{-.25em}{\rm C} \else \mbox{$\hspace{.2em}\vbarind\hspace{-.25em}{\rm C}$} \fi}
\newcommand{\Q}{\ifmmode \hspace{.2em}\vbar\hspace{-.31em}{\rm Q} \else \mbox{$\hspace{.2em}\vbar\hspace{-.31em}{\rm Q}$} \fi}
\newcommand{\Z}{\ifmmode {\rm Z}\hspace{-.28em}{\rm Z} \else ${\rm Z}\hspace{-.28em}{\rm Z}$ \fi}

\renewcommand{\vec}[1]{\bf{#1}}     

